\documentclass[compsoc,conference,a4paper,10pt,times]{IEEEtran}
\IEEEoverridecommandlockouts

\usepackage{tikz}
\usepackage{amsmath}

\usepackage{amssymb}
\usepackage{eucal}
\usepackage{filecontents}
\usepackage{epsfig,endnotes}
\usepackage{xspace}
\usepackage{amsmath}
\usepackage{xcolor}
\usepackage{pifont}
\usepackage{todonotes}
\usepackage{hyperref}
\usepackage[normalem]{ulem}
\usepackage{balance}
\usepackage{gensymb}
\usepackage{multirow}
\usepackage{subcaption}
\usepackage{caption}
\usepackage{mathtools}
\usepackage{bbm}
\usepackage{pifont}
\usepackage{balance}
\usepackage[keeplastbox]{flushend}

\newcommand{\etal}{\textit{et al}.}
\newcommand{\eg}{e.g., \xspace}
\newcommand{\ie}{i.e., \xspace}
\def\approach{{\sc ANDRuspex}\xspace}

\usepackage{amsthm}
\theoremstyle{definition}

\begin{document}

\title{\approach: Leveraging Graph Representation Learning to Predict Harmful App Installations on Mobile Devices}

\author{\IEEEauthorblockN{Yun Shen}
\IEEEauthorblockA{
\textit{NortonLifeLock Research Group}\\
yun.shen@nortonlifelock.com}
\and
\IEEEauthorblockN{Gianluca Stringhini}
\IEEEauthorblockA{
\textit{Boston University}\\
gian@bu.edu}
}

\maketitle

\begin{abstract}
Android's security model severely limits the capabilities of anti-malware software.
Unlike commodity anti-malware solutions on desktop systems, their Android counterparts run as sandboxed applications without root privileges and are limited by Android's permission system.
As such, PHAs on Android are usually willingly installed by victims, as they come disguised as useful applications with hidden malicious functionality, and are encountered on mobile app stores as suggestions based on the apps that a user previously installed. 
Users with similar interests and app installation history are likely to be exposed and to decide to install the same PHA.
This observation gives us the opportunity to develop predictive approaches that can warn the user about which PHAs they will encounter and potentially be tempted to install in the near future.
These approaches could then be used to complement commodity anti-malware solutions, which are focused on post-fact detection, closing the window of opportunity that existing solutions suffer from.
In this paper we develop \approach, a system based on graph representation learning, allowing us to learn latent relationships between user devices and PHAs and leverage them for prediction.
We test \approach on a real world dataset of PHA installations collected by a security company, and show that our approach achieves very high prediction results (up to 0.994 TPR at 0.0001 FPR), while at the same time outperforming alternative baseline methods.
We also demonstrate that \approach is robust and its runtime performance is acceptable for a real world deployment.
\end{abstract}

\section{Introduction}
\label{sec:introduction}

While becoming the most popular mobile OS in the world, Android has also attracted significant interest by malicious parties, who started developing apps with malicious purposes targeting this platform~\cite{allix2016androzoo,lever2013core,suarez2018eight,zhou2012dissecting}. 
Millions of these malicious Android apps are observed every year~\cite{allix2016androzoo}, carrying out various types of harmful activity including stealing private information from the victim devices~\cite{shen2021understanding}, sending premium SMS messages, performing click fraud, and encrypting the victim's data in exchange for a ransom~\cite{andronio2015heldroid,mirzaei2019andrensemble}. 
Similar to what happens for desktop computers, not all malicious Android applications come with clearly harmful content, but some present unwanted components that are often an annoyance to the user (\eg adware~\cite{erturk2012case,ife2019waves,kotzias2016measuring,thomas2016investigating}) or without user consent (\eg click fraud~\cite{crussell2014madfraud}).
In this paper, we adopt the terminology used by Google and refer to these potentially unwanted Android apps as \emph{potentially harmful apps} (PHAs)~\cite{googlereport}.
Unlike traditional desktop malware, which is often installed by automatically exploiting vulnerabilities in the victim's Web browser and carrying out \emph{drive-by} download attacks~\cite{provos2007ghost}, Android PHAs tend to be manually installed by victims, who willingly choose to install apps that promise useful functionalities but come together with harmful code or advertising SDKs that hide malicious functionalities~\cite{suarez2018eight,wermke2018large,zhou2012dissecting}. 
In fact, the preliminary analysis that we performed ahead of this study on a real world dataset of apps installed by millions of real users showed that 93.1\% of PHAs are manually installed via either the Google play store or other well known side-loading apps like \texttt{com.sec.android.easyMover} and \texttt{com.huawei.appmarket}.

To mitigate the threat of PHAs on Android, the security community has followed two directions.
The first one is to analyze Android apps to identify malicious behavior~\cite{aafer2013droidapiminer,arp2014drebin,grace2012riskranker,peng2012using,chakradeo2013mast,mariconti2016mamadroid,yang2014droidminer,yuan2014droid}.
In the real world, this analysis usually takes place when apps are submitted to an Android marketplace. 
For example, Google uses a vetting system known as \emph{Android Bouncer}~\cite{oberheide2012dissecting}. 
While quite effective in detecting malicious Android apps, these systems suffer from the limitation that the Android ecosystem is quite fragmented, and users install apps from a variety of sources, including alternative marketplaces that do not employ strict enough security measures~\cite{allix2016androzoo,wang2018beyond}.
To complement market-level detection, a plethora of commercial anti-malware software products are available to users. 
Unfortunately, the Android security model severely limits their capabilities~\cite{enck2009understanding}.
Unlike traditional desktop anti-malware software, mobile solutions run as sandboxed apps without root privileges~\cite{enck2009understanding} and cannot proactively block malicious downloads or even automatically remove a PHA once they find it; instead, they periodically scan the mobile device for suspicious apps and simply warn the user about newly discovered threats, prompting them to manually remove the detected PHAs. 
This is far from ideal because detection happens after the PHAs have been installed, leaving a \emph{window of vulnerability} during which attackers can cause harm to the victims and their devices undetected. 

In this paper, we propose to mitigate the aforementioned issues by complementing existing on-device anti-malware solutions with the capability of \emph{predicting} future PHA installations on the device.
Our intuition is that since the installation of a PHA on Android is usually a consequence of a user's own rational choices, the installation of a certain PHA on a user's device can be predicted by observing the apps that the user has installed in the past, together with which PHAs ``similar'' users have installed.
For example, users with similar interests (\eg gaming) might be active on the same third party marketplaces and receive suggestions to install similar PHAs (\eg disguised as free versions of popular games).
If we were able to predict which PHA a user will attempt to install in the near future, we could display a warning to them ahead of time and potentially convince them not to install that app, closing the window of vulnerability between installation and detection introduced by traditional Android anti-malware software.

To achieve an effective prediction of which PHAs will be installed on mobile devices in the future, we develop a system called \approach.
Our approach is based on \emph{graph representation learning}~\cite{hamilton2017representation}.
This technique allows us to automatically discover important features in raw data, without the need for feature engineering.
\approach first builds a graph of PHA installation events on a global scale, in which edges represent which devices installed which PHAs.
It then applies representation learning to learn the low dimensional vertex representation of the PHA installation graph at time $t$.
Finally, \approach predicts new links that will be formed between devices and PHAs that at time $t+\delta$ based on their respective properties and the currently observed links at time $t$.
In production, \approach would then warn the user about suspicious applications that they will encounter and likely be interested in installing in the near future (\eg on a third party Android marketplace), complementing current on device Android anti-malware solutions that can only act after the fact.

\noindent \textbf{Contributions.} Our contributions are summarized as follows;
\begin{itemize}
\item We present \approach, a system based on graph representation learning which allows us to predict which PHAs a user might be tempted to install on an Android device ahead of time.
 To the best of our knowledge, we are the first ones to propose a predictive system in the area of Android malware mitigation.
\item We test \approach on a dataset of tens of millions of PHA installations collected from real devices that installed the Android anti-malware program of a major security company.
We show that \approach can achieve very good performance prediction (up to 0.994 TPR at 0.0001 FPR), significantly outperforming simpler baseline methods. 
\item We also show that \approach could be effectively used in the wild since it has both acceptable performance and it does not significantly reduce its detection performance when the system needs to be retrained.
\end{itemize}

\section{Motivation}
\label{sec:motivation}

\begin{figure}[t]
    \centering
    \includegraphics[width=\linewidth]{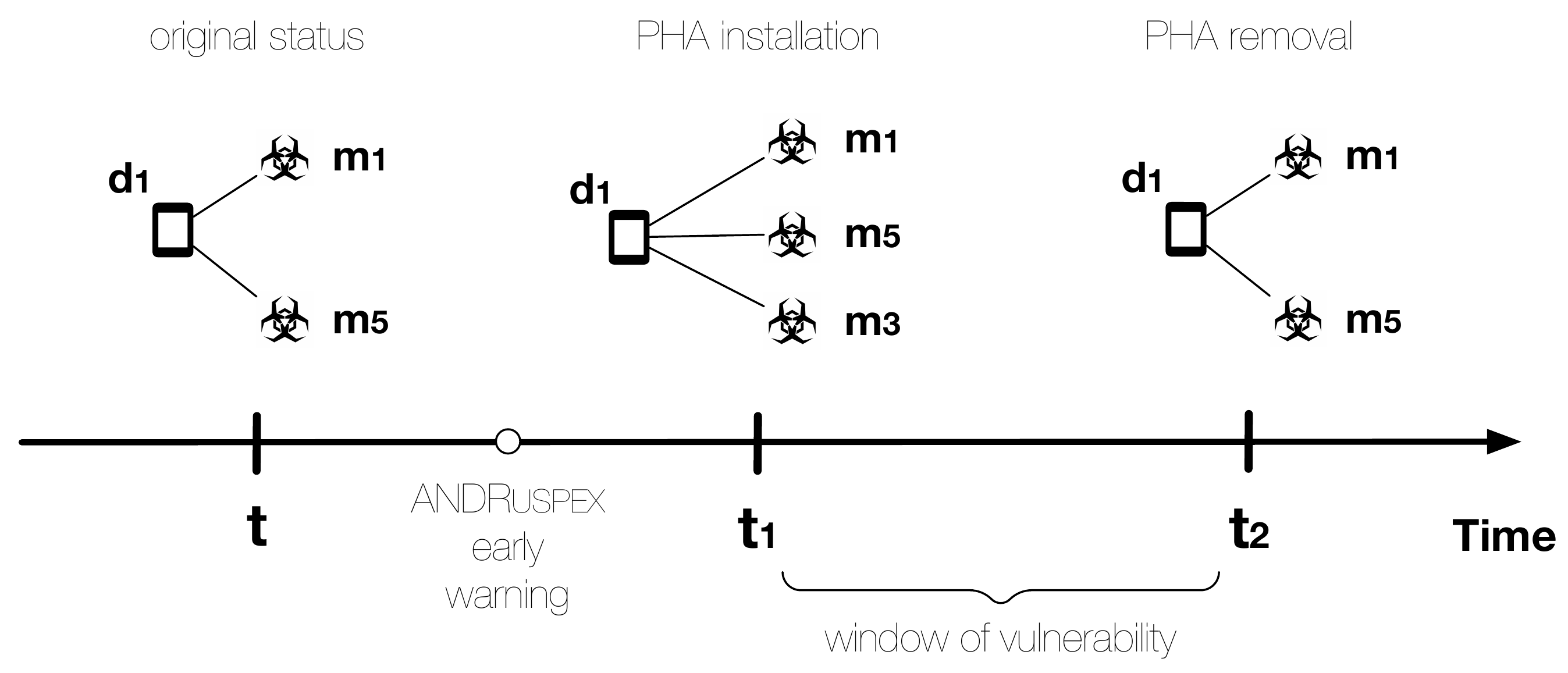}
    \caption{PHA installation over time for a device $d_1$. The goal of our approach is to correctly predict that $d_1$ will install $m_3$ in the future and close the window of vulnerability.}
    \label{fig:example}
\end{figure}

\begin{figure*}[t]
     \centering
     \resizebox{0.9\linewidth}{!} {
     \begin{subfigure}[t]{0.55\textwidth}
         \includegraphics[width=\linewidth]{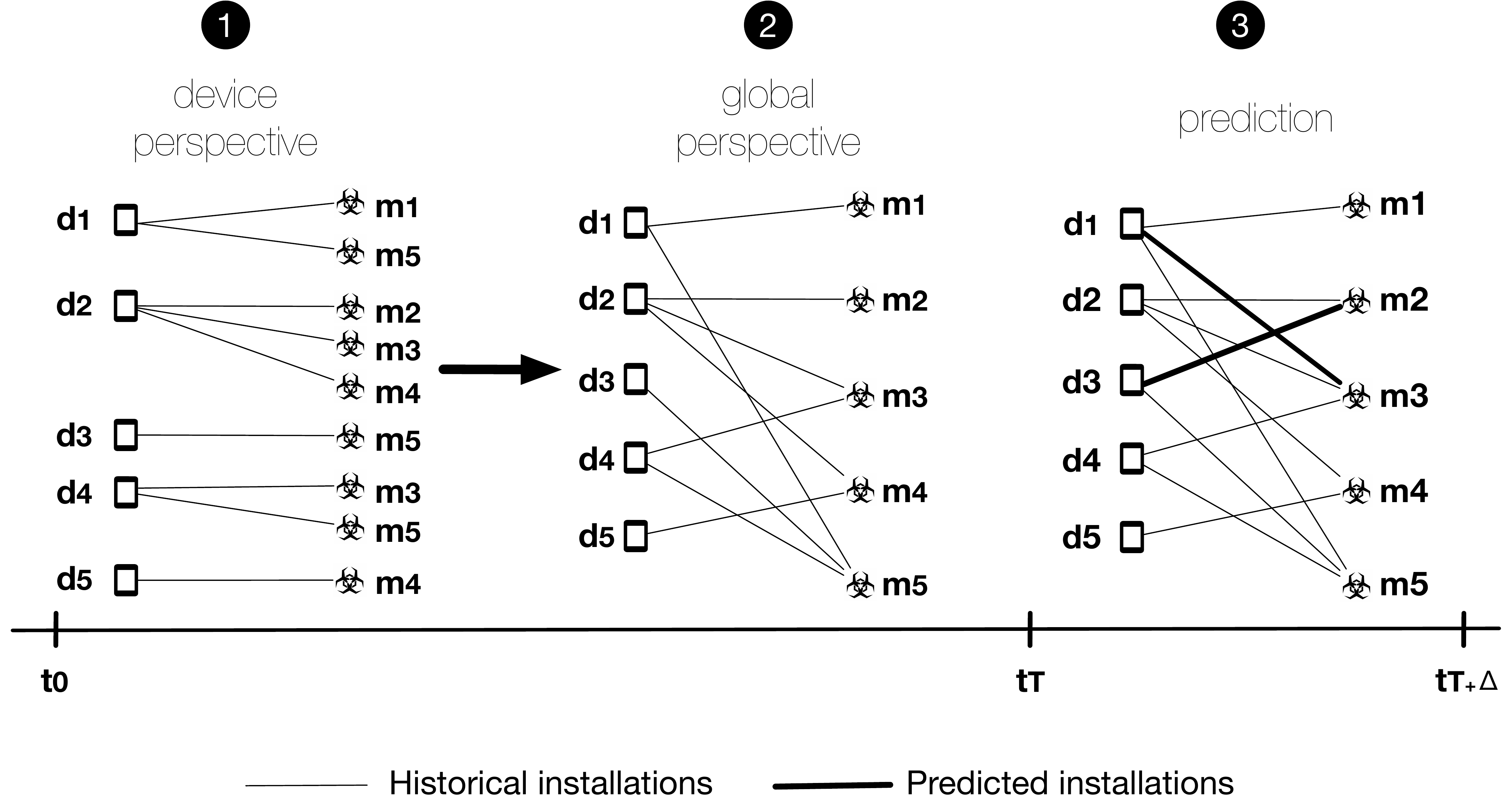}
         \caption{}
         \label{fig:motivation}
     \end{subfigure}
     \hfill
    \begin{subfigure}[t]{0.38\textwidth}
        \includegraphics[width=\linewidth]{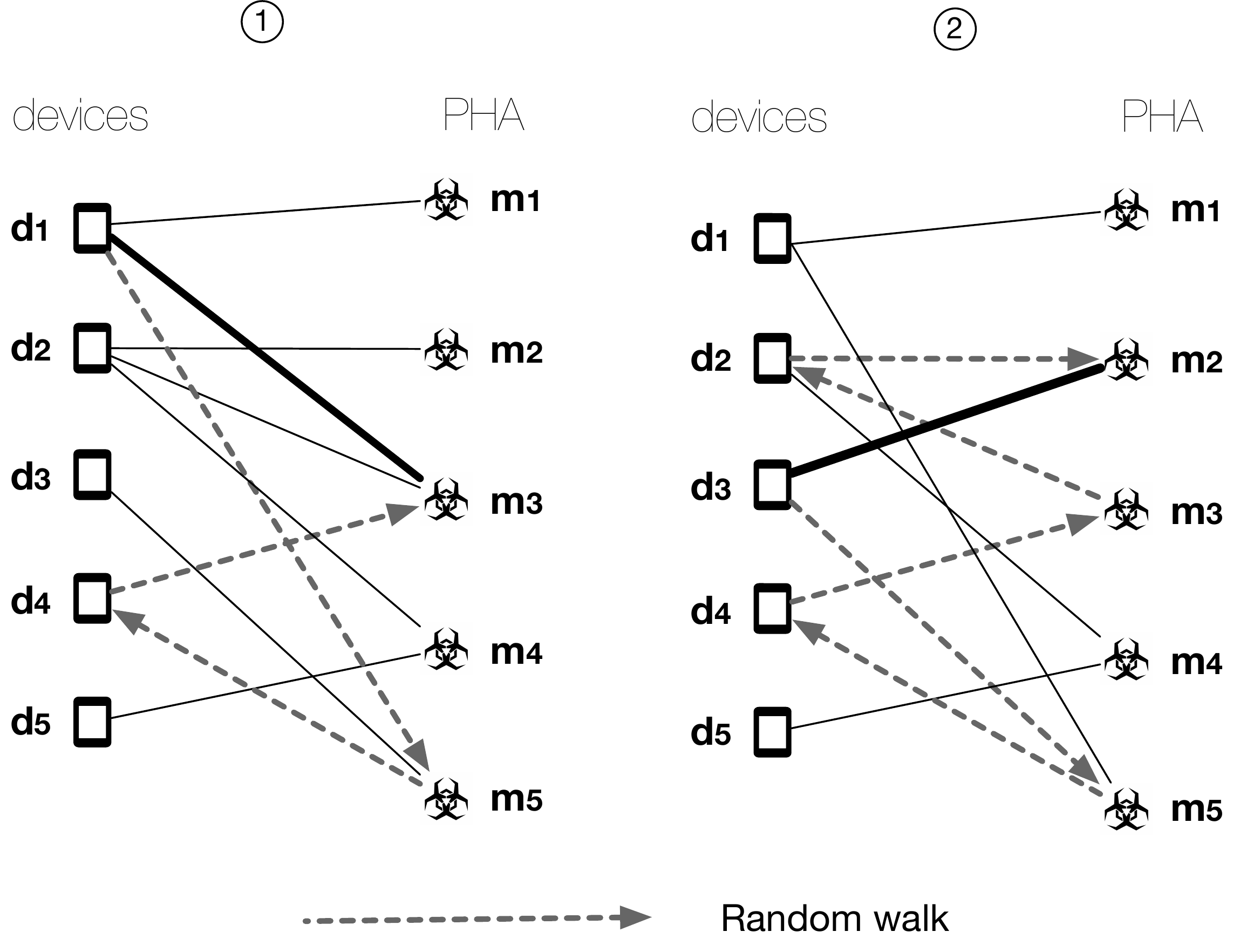}
        \caption{}
        \label{fig:rw_prediction}
    \end{subfigure}
    }
    \hfill
    \caption{Global view of PHA installation events (Figure~\ref{fig:motivation}) and potential prediction clues captured by random walks (Figure~\ref{fig:rw_prediction}) from this global graph.}
    \label{fig:prediction_examples}
\end{figure*}

\begin{figure}[t]
     \centering
     
    \begin{subfigure}[t]{0.4\textwidth}
        \includegraphics[width=\linewidth]{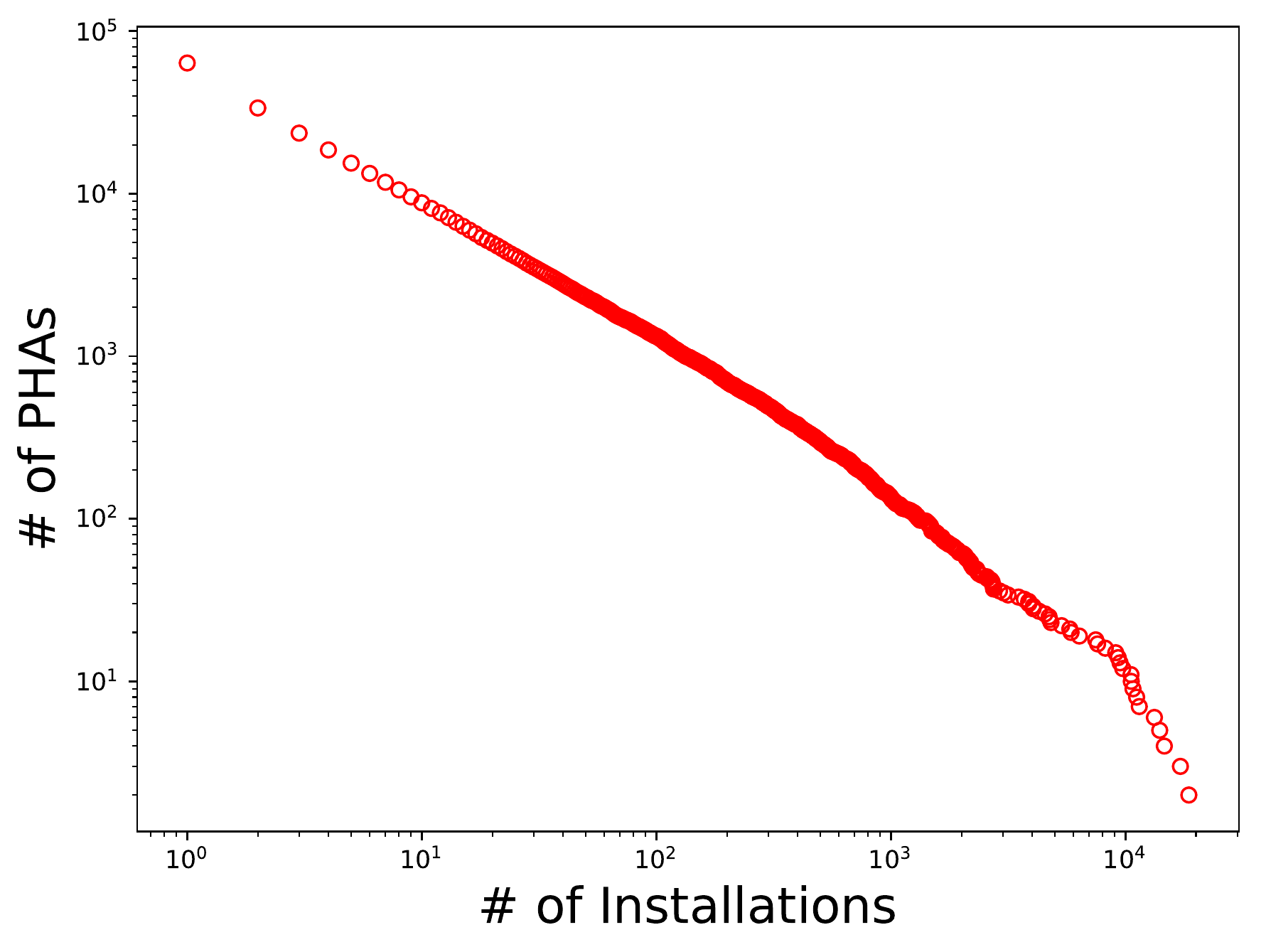}
        \caption{}
        \label{fig:malware_degree_distribution}
    \end{subfigure}
    \hfill
    \begin{subfigure}[t]{0.4\textwidth}
        \includegraphics[width=\linewidth]{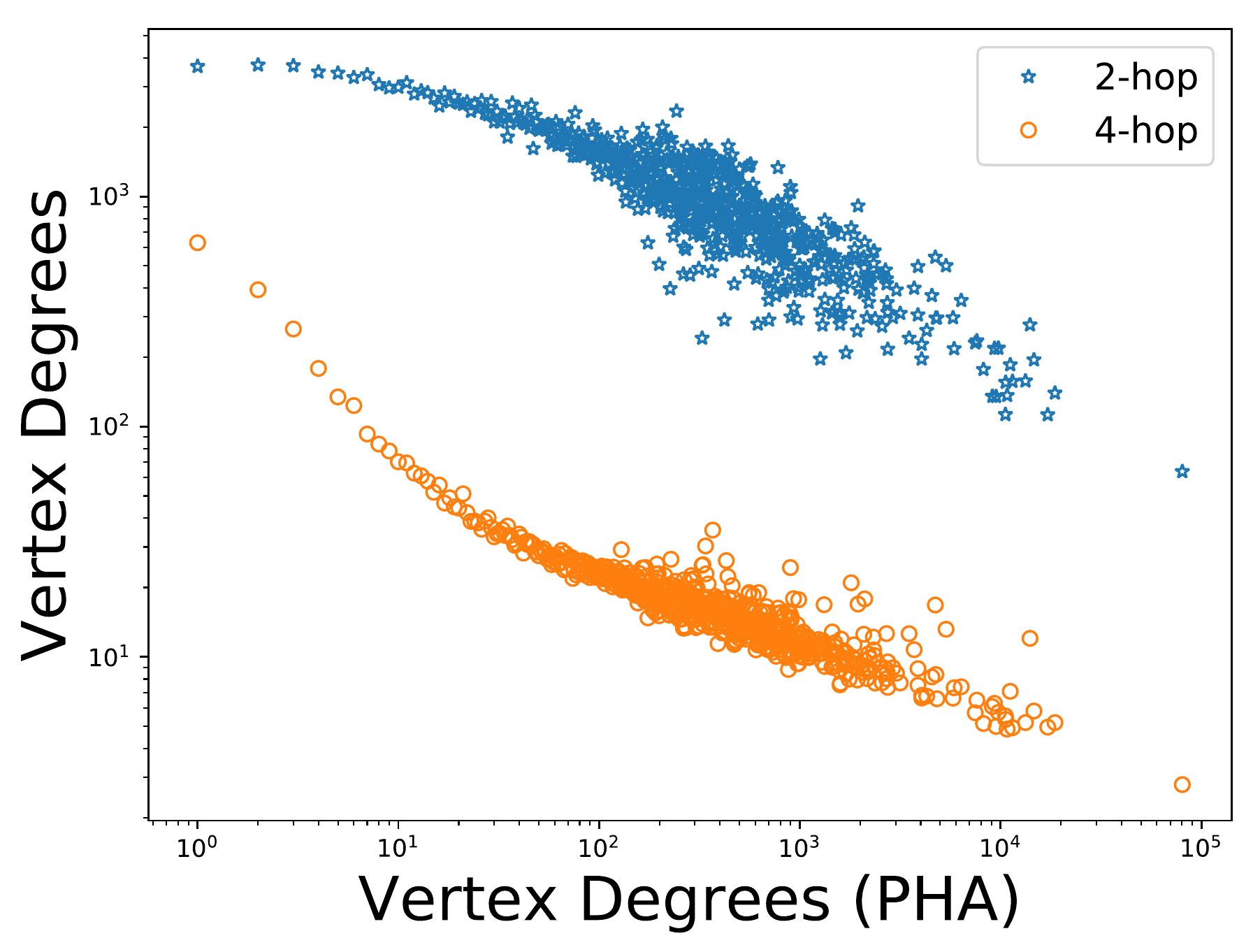}
        \caption{}
        \label{fig:neighbour_of_neighbour}
    \end{subfigure}
    
    \hfill
    \caption{PHA installations follow a power law distribution (Figure~\ref{fig:malware_degree_distribution}) and have implicit correlations (Figure~\ref{fig:neighbour_of_neighbour}) between PHA degrees and the average degrees of all vertices reachable by 2 hops (\ie by a path ($\mathbf{m}_{i}, \mathbf{d}_{z}, \mathbf{m}_{j}$), \textcolor{blue}{blue}) and 4 hops (\ie by a path ($\mathbf{m}_{i}, \mathbf{d}_{z}, \mathbf{m}_{j}, \mathbf{d}_{l}, \mathbf{m}_{h}$), \textcolor{orange}{orange}) using one day PHA installation data (see Section~\ref{sec:dataset}). }
    \label{fig:degree_distribution}
\end{figure}

The main motivation behind our approach is that, unlike desktop malware, PHAs on Android are usually willingly installed by less attentive users~\cite{felt2012android}.
To confirm this motivation, we performed a preliminary analysis of the PHA installation events reported by 2.4M devices that installed a commercial mobile security product during March 2019 (see Section~\ref{sec:dataset} for more details about the dataset).
When a PHA is detected, the security product records its installer package names and how it was installed on the device (\ie pre-installed or user-installed). 
Matching this information with the official package names of Android market apps we find that 93.1\% of the PHAs identified on those devices were manually installed by users through Google Play or alternative marketplaces. 
This is confirmed by a recent work by Kotzias~\etal~\cite{kotzias2021did}, which showed that the Google Play market is responsible for 87\% of all PHA installations and for 67\% of unwanted app installations.
Since the Google Play market is the official market of Android apps,
end users may trust apps distributed from Google Play and willingly install them.
In short, our preliminary analysis, combined with the fact that anti-malware systems on Android are severely limited by the platform's security model, and cannot proactively block known PHAs nor automatically remove them after they are detected, shows that a predictive approach to PHA mitigation, allowing the user to be warned ahead of time, would be beneficial.

The main research challenge is how to do this effectively.
Looking at device activity in isolation does not provide enough information to predict the upcoming PHA installations for a given mobile device.
For example, in Figure~\ref{fig:example} there is no easy way to predict that $m_3$ will be installed based on the past installation activity on $d_1$.
One possible solution is to collect abundant additional information  about $d_1$ (\eg application categories, device types, marketplace information, etc.) and leverage feature engineering to create features for PHA installation events. 
These features can then be used to train a machine learning model to predict future installations. However, such process requires domain knowledge and is essentially a manual endeavor that is time-consuming and problem-dependent.

Another plausible solution is looking at the installation behavior of multiple devices, treating a device's historical installation activities as its behavioral profile, and grouping devices with similar profiles together~\cite{zhang2008highly}.  
This way, we can predict which PHA a given device's owner will install based upon the PHAs that have been installed by those with similar profiles. 
For instance, as we can see in Figure~\ref{fig:motivation} (\ding{182}), $\mathbf{d}_1$, $\mathbf{d}_3$, and $\mathbf{d}_4$ share similar behavioral profiles. 
Given how Android marketplaces work, it is likely that $\mathbf{m}_1$ and $\mathbf{m}_3$ are apps that the owner of $\mathbf{d}_3$ is interested in, and they might even been suggested by the marketplace itself. 
As such, it is more likely that $\mathbf{d}_3$ will install those PHAs in the future than a randomly chosen PHA.
%It is more likely for $\mathbf{d}_3$ to install either $\mathbf{m}_1$ and $\mathbf{m}_3$ in the future rather than a randomly chosen PHA. 
It is however not trivial to group devices with similar behavioral profiles. For example, we can use the Jaccard similarity coefficient to calculate the similarity between two devices, but this leads to a $O(n^2)$ time complexity method~\cite{zhang2008highly} to generate pair wise similarity for potentially millions of devices, which renders such approach practically infeasible.

To mitigate the aforementioned limitations, we propose to approach the problem of predicting PHA installations in a principled way, by looking at the collective PHA installation events from a global perspective instead of focusing on a device level (see \ding{183} in Figure~\ref{fig:motivation}). As it can be seen, this graph provides aggregated historical information of how the PHAs have been installed by mobile devices globally, and yields additional structural information by representing the relations between devices and PHAs as a bipartite graph. This graph enables us to capture the collective behavior from all devices, and elicit the similarities between the download activity of different devices. 

Leveraging this global installation graph for prediction, however, presents two challenges. 
The first challenge is the distribution of PHA installations among the devices. We use data on the PHAs installed by real devices over a period of one day (see Section~\ref{sec:dataset}) for illustration purposes. As we can see in Figure~\ref{fig:malware_degree_distribution}, the distribution of PHA installations (\ie vertex degrees from a graph perspective) follows a power law distribution. 
This indicates that there are popular PHAs with a number of installations that greatly exceed the average and that the majority of the PHA population only appears in a small number of devices. Preferential attachment~\cite{barabasi2002evolution,newman2001clustering} has been proposed as an effective mechanism to explain conjectured power law degree distributions. We show in Section~\ref{sec:exp_comparison} that such model is promising but insufficient to predict PHA installation, which instead requires a model that considers both popular and less popular PHAs at the same time.

The second challenge is modeling the implicit relationships among PHAs with limited information. 
For example, in the real world, a user may choose to install $\mathbf{m}_1$ due to a deceitful full-screen ad displayed in $\mathbf{m}_5$, and $\mathbf{m}_3$ could be willingly installed by a user on device $\mathbf{d}_4$ because recommended by the same app store from which $\mathbf{m}_5$ was initially installed. 
Inferring these relationships from observing app installation is challenging.
The global installation graph as depicted in Figure~\ref{fig:motivation} \ding{184} can help identify these hidden relationships.
In Figure~\ref{fig:neighbour_of_neighbour}, we show possible correlations between PHA degrees and the average degrees of all vertices reachable by 2 hops (\ie by a path ($\mathbf{m}_{i}, \mathbf{d}_{z}, \mathbf{m}_{j}$), blue) and 4 hops (\ie by a path ($\mathbf{m}_{i}, \mathbf{d}_{z}, \mathbf{m}_{j}, \mathbf{d}_{l}, \mathbf{m}_{h}$), yellow). We can observe two interesting aspects. The first aspect is that there is a negative correlation in both cases. This indicates that PHAs with larger installations (\ie popular PHAs) are co-existing with smaller ones (\ie less popular PHAs). The second aspect is that the correlation coefficient decreases with the increasing number of hops. For example, the correlation coefficient between PHA degrees and the average degrees of 2-hop vertices is $-0.28$ while the correlation coefficient of those 4 hops away is $-0.11$.

\noindent \textbf{Core idea.} To address both challenges, in this paper, we put random walk~\cite{weiss1994aspects,norris1998markov} and graph representation learning~\cite{bengio2013representation} in use. 
Graph representation learning allows us to take both popular and unpopular PHAs into account, while random walk enables us to harvest the aforementioned explicit and implicit relationships among PHAs for each device. 
Take Figure~\ref{fig:rw_prediction} (\ding{192}) for instance, if we want to predict the next PHA that  $\mathbf{d}_1$ will install, we can take a random walk starting from $\mathbf{d}_1$ to capture the aforementioned implicit relationship. Suppose we obtain a random walk $(\mathbf{d}_{1}, \mathbf{m}_{5}, \mathbf{d}_{4}, \mathbf{m}_{3})$ as illustrated in Figure~\ref{fig:rw_prediction} (\ding{192}), offers us an indication that $\mathbf{d}_{1}$ may potentially install the malicious app $\mathbf{m}_{3}$ in the near future. Similarly, as we can see  in Figure~\ref{fig:rw_prediction} (\ding{193}), a random walk $(\mathbf{d}_{3}, \mathbf{m}_{5}, \mathbf{d}_{4}, \mathbf{m}_{3}, \mathbf{d}_{2}, \mathbf{m}_{2})$ may capture the higher proximity relationship between $\mathbf{d}_{3}$ and $\mathbf{m}_{2}$, and in turn, facilitate the prediction. In Section~\ref{sec:methodology} we show that leveraging graph representation learning to extract useful information from the random walks enables us to understand the explicit and implicit dynamics between devices and PHAs without a time-consuming feature engineering step, and allows us to make accurate PHA installation predictions up to one week ahead of the real installations.

\noindent \textbf{Problem formulation.} We formalize our PHA installation prediction problem as follows. Let $\mathbf{D}$ and $\mathbf{M}$ denote all unique devices and PHAs respectively, and $\mathbf{E}$ denote all observed PHA installation events. We define our global PHA installation graph as a bipartite graph $\mathbf{G}_{[t_0, t_T]} = (\mathbf{V}, \mathbf{E})$, where $\mathbf{V} = \mathbf{D} \cup \mathbf{M}$ denotes vertices in the graph, each edge $\mathbf{e}=(\mathbf{d_i}, \mathbf{m_j}, t_k) \in \mathbf{E}$ represents a device $\mathbf{d_i} \in \mathbf{D}$ installs a PHA $\mathbf{m_j} \in \mathbf{M}$ at a particular timestamp $t_k \leq t_T$. Our goal is to learn lower $d$-dimensional representations of $\mathbf{V}$ in $\mathcal{G}_{[t_0, t_T]}$, denoted as $\Phi_\mathbf{V} \in \mathbb{R}^{|\mathbf{V}| \times d}$, by taking global PHA installation information into consideration. With the learned vertex embeddings $\Phi_\mathbf{V}$, \approach trains a prediction function $f: \mathbf{G}_{(t_0, t_{T}]} \rightarrow \mathbf{E}_{(t_T, t_{T+\Delta}]}$ that outputs a list of edges (\ie PHA installations) that are not present in $\mathbf{G}_{[t_0, t_T]}$, but predicted to appear in the graph $\mathbf{G}_{(t_T, t_{T+\Delta}]}$ (see \ding{184} in Figure~\ref{fig:motivation} for illustration).

\noindent \textbf{Note.} In this paper, we do not aim to understand how PHAs are distributed nor do we target to detect unknown PHAs through graph analysis. We also do not investigate, from a behavioral perspective, how PHAs operate and propagate once installed in the devices. Rather, our main goal is to predict the exact PHAs that are likely to be installed by the end users \textbf{in the near future} given their historical installation activities and the collective PHA installation events on a global scale.

\section{Methodology}
\label{sec:methodology}

\subsection{System Overview}

\begin{figure}[t]
    \centering
    \includegraphics[width=0.9\linewidth]{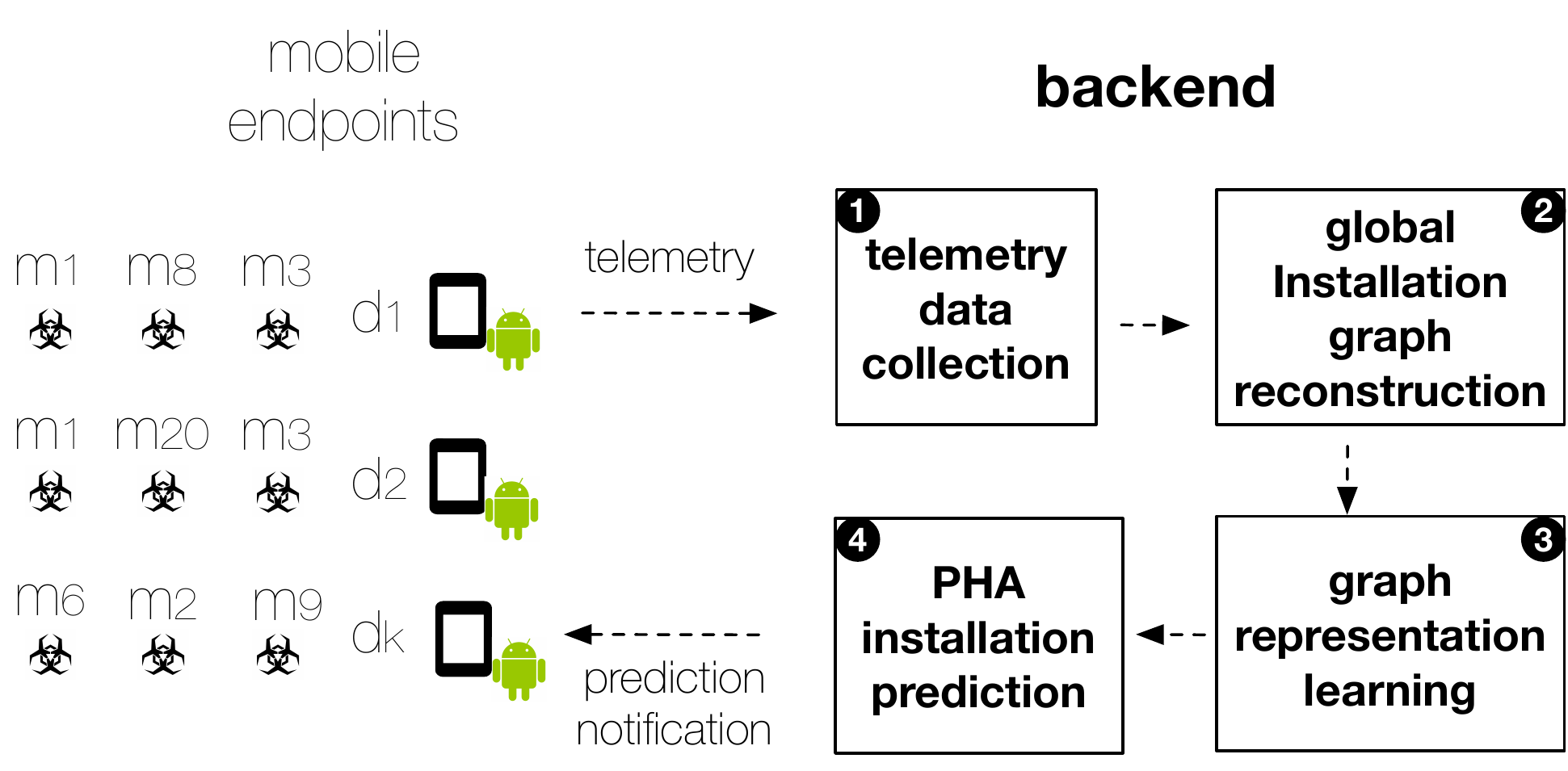}
    \caption{System overview of \approach}
    \label{fig:system_overview}
\end{figure}

The system overview of \approach is depicted in Figure~\ref{fig:system_overview}. 
\approach's frontend component is installed on mobile devices, alongside an existing mobile malware detection product.
Every time the malware detection product detects that a PHA has been installed, the mobile device sends data to our backend infrastructure (\ding{182}). 
\approach's backend component uses this data to build a global PHA installation graph (\ding{183}). 
It then leverages graph representation learning (Section~\ref{sec:embdding}) to capture the collective behavior from all devices and understand the implicit relationships among the PHAs in a latent space (\ding{184}). 
\approach periodically trains a prediction engine based upon these historical installations, and predicts the impending PHA installations on the mobile devices that installed the frontend component in the future (\ding{185}). 
When \approach predicts that a mobile device's user will attempt to install a certain PHA in the near future (for example because this PHA will appear as a suggestion on a third party market) it sends an alert to the mobile device, so that the device's owner can be warned about the threat and reach an informed decision to not install a PHA which might weaken their device security. 
\label{sec:system_overview}

\subsection{\approach Workflow}

\begin{figure*}[t]
    \centering
    \includegraphics[width=0.8\linewidth]{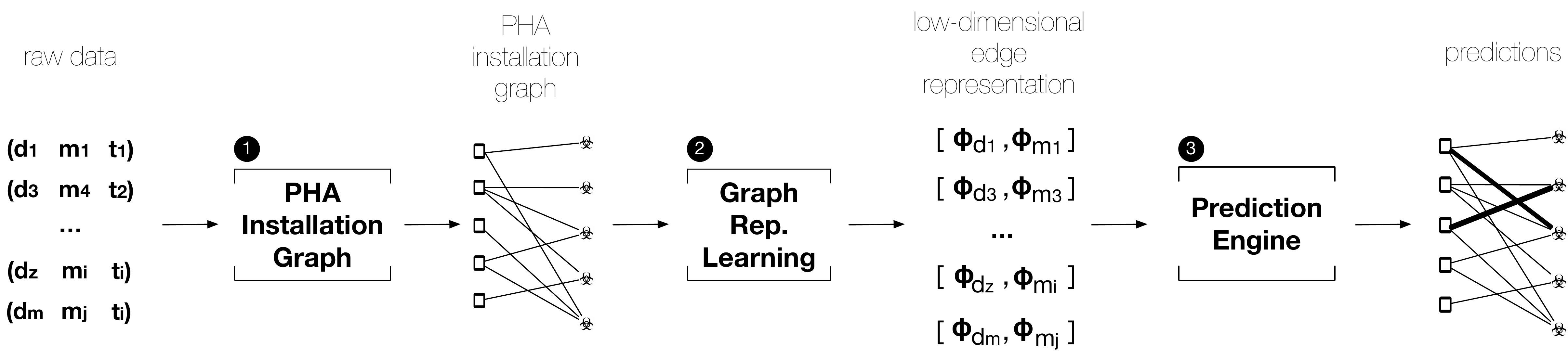}
    \caption{\approach's architecture. \approach collects PHA installation events from the mobile devices that have installed a mobile security product. These events are used to build a global PHA installation graph. \approach then learns the low dimensional vertex representations and uses them to build and validate \approach's predictive model. }
    \label{fig:architecture}
\end{figure*}

Following the system overview in Figure~\ref{fig:system_overview}, the workflow of \approach at the backend is depicted in Figure~\ref{fig:architecture}. Its operation consists of three steps: \ding{182} reconstruction of the global PHA installation graph, \ding{183} graph representation learning, and \ding{184} PHA installation prediction. 

\noindent \textbf{Step 1: Reconstruction of the Global PHA installation graph (\ding{182} in Figure~\ref{fig:architecture}).} The goal of this step is building an installation graph that encapsulates a comprehensive view of how PHAs are installed by mobile devices on a global scale. To build the graph, \approach takes as input the streams of detected PHA installation events generated by the mobile devices. These events are treated as tuples, in the format of $(\mathbf{d}_i, \mathbf{m}_j, t_k)$. Let $\mathbf{A}$ denote the symmetric adjacency matrix of $\mathbf{G}$. $\mathbf{A}_{\mathbf{d}_i, \mathbf{m}_j} = 1$ if $\mathbf{d}_i$ and $\mathbf{m}_j$ are linked in the same installation event, otherwise $\mathbf{A}_{\mathbf{d}_i, \mathbf{m}_j} = 0$. The output of this step is a global PHA installation graph $\mathbf{G}$ represented by the adjacency matrix $\mathbf{A}^{|\mathbf{D}| \times |\mathbf{M}|}$. This matrix will be used by the graph representation learning module in the next step.

\noindent \textbf{Step 2: Graph representation learning (\ding{183} in Figure~\ref{fig:architecture}).} 
The core of \approach is learning the low dimensional vertex representations (\ie vertex embeddings) of the PHA installation graph. 
Essentially, \approach takes a number of truncated random walks from each vertex $\mathbf{d}_i \in \mathbf{D}$ in the graph $\mathbf{G}$. These random walks effectively capture the high-order indirect connections between mobile devices and PHAs. 
In this way, \approach explicitly builds a high-order proximity transition matrix between $\mathbf{D}$ and $\mathbf{M}$. 
It then factorizes this matrix together with a decay factor to account for the strength of the implicit connections (see technical details in Section~\ref{sec:embdding}). 
The output of this step is the low dimensional vertex representations (\ie $\Phi_{\mathbf{V}}$) that will be used to model PHA installation events in the latent space.

\noindent \textbf{Step 3: PHA installation prediction (\ding{184} in Figure~\ref{fig:architecture}).} 
Taking the vertex embeddings as the input, \approach models the observed PHA installation events in the latent space and trains a prediction model to predict future installations. 
First, \approach models the observed PHA installations by concatenating two low dimensional vertex embeddings in the latent space (see Section~\ref{sec:case_study} for the influence of alternative edge representations). 
That is, $\phi_{\mathbf{e}} = concat(\phi_{\mathbf{d_i}}, \phi_{\mathbf{m_j}})$, where $\mathbf{e} \in \mathbf{E}$. 
Effectively, each PHA installation event is represented by a $2d$-dimensional vector. 
\approach then formulates the prediction task as a binary prediction problem for edges (\ie PHA installations) in the graph where two classes are considered: \emph{positive} or presence of edges and \emph{negative} or absence of edges. 
\approach samples an equal number of non-existing edges~\cite{pendlebury2019tesseract} from the observed PHA installation edges (\ie positive class) to form the negative class and train the prediction model. 
The output of the PHA installation prediction is a list of edges (\ie PHA installations) that are not present in $[t_0, t_T]$, but predicted to appear in the $(t_T, t_{T+\Delta}]$.
These predictions can then be leveraged to warn the user ahead of time about PHAs that they are likely to encounter and might be enticed to install.

\noindent \textbf{End-to-end solution.}
We envision \approach to be deployed as an end-to-end solution to warn the end user about PHAs that they will likely encounter in the future.
At the backend, \approach continuously collects and monitors PHAs installed on all mobile devices that install the security application.
If an end user installs a known PHA, the installation event would trigger the on-device security engine to send the detection back to \approach.
The system would then calculate the probability of this end user to install a related PHA based on the telemetry from other devices.
If the probability is higher than a threshold, \approach notifies the end user with a summary of why the system reached such a decision.
For example, \approach could highlight the key aspects of a PHA distribution campaign as shown in the case studies in Section~\ref{sec:case_study}.
This summary would be displayed to the end user when they use the security app as in-app messages.
The rationale of using in-app messages is to explain to the end user the potential risk associated with the known on-device PHAs.
This way, the end user would benefit from an informative explanation rather than a short notification via the status bar.
We are planning to collaborate with usable security experts in the future to design an effective in-app warning approach(\eg modularity based approach ~\cite{micallef2017stop}.

\subsection{Graph Representation Learning}
\label{sec:embdding}

\begin{figure}[t]
     \centering
    \begin{subfigure}[t]{0.22\textwidth}
        \includegraphics[width=\linewidth]{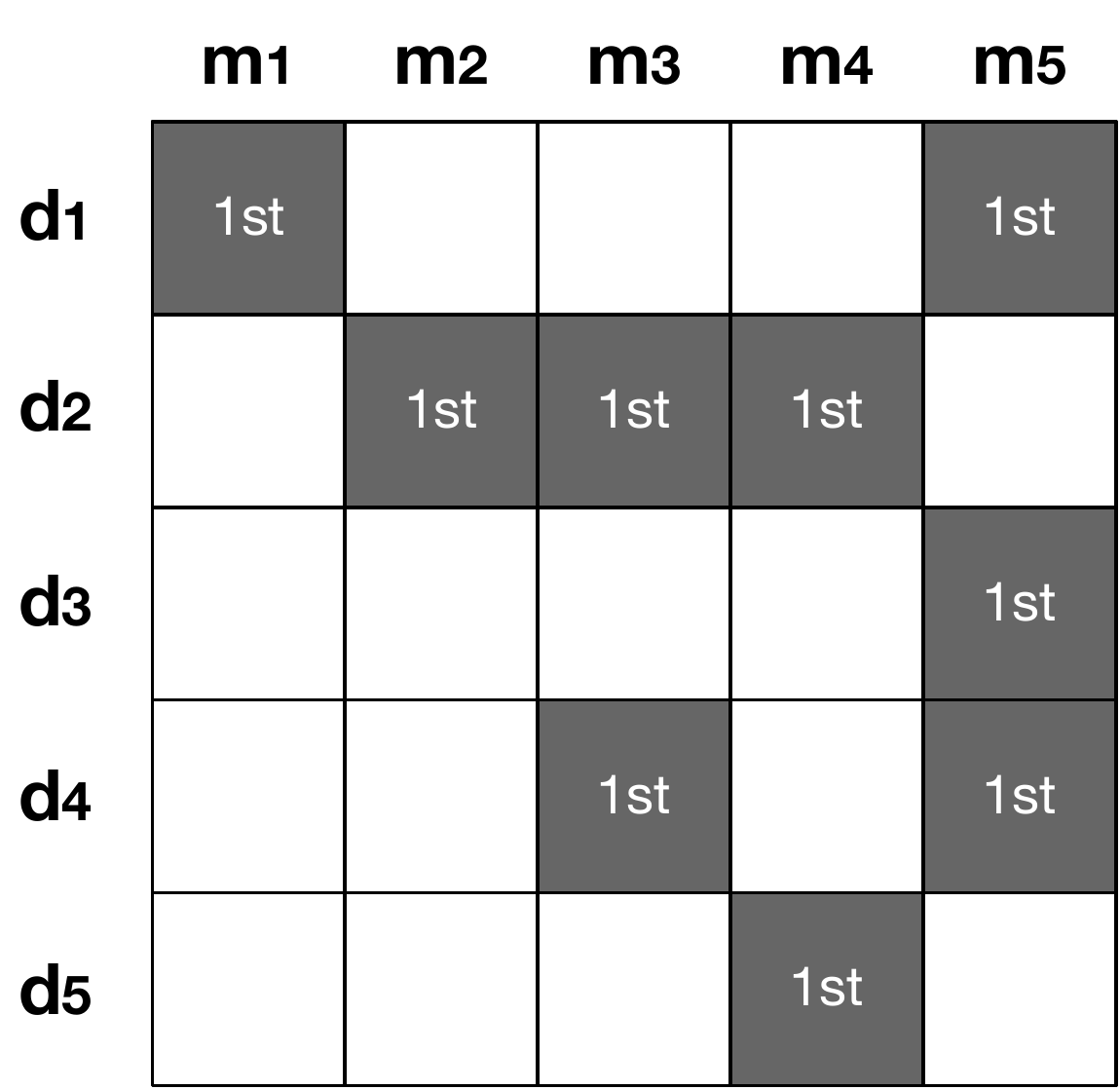}
        \caption{}
        \label{fig:first_order_matrix}
    \end{subfigure}
    \hfill
    \begin{subfigure}[t]{0.22\textwidth}
        \includegraphics[width=\linewidth]{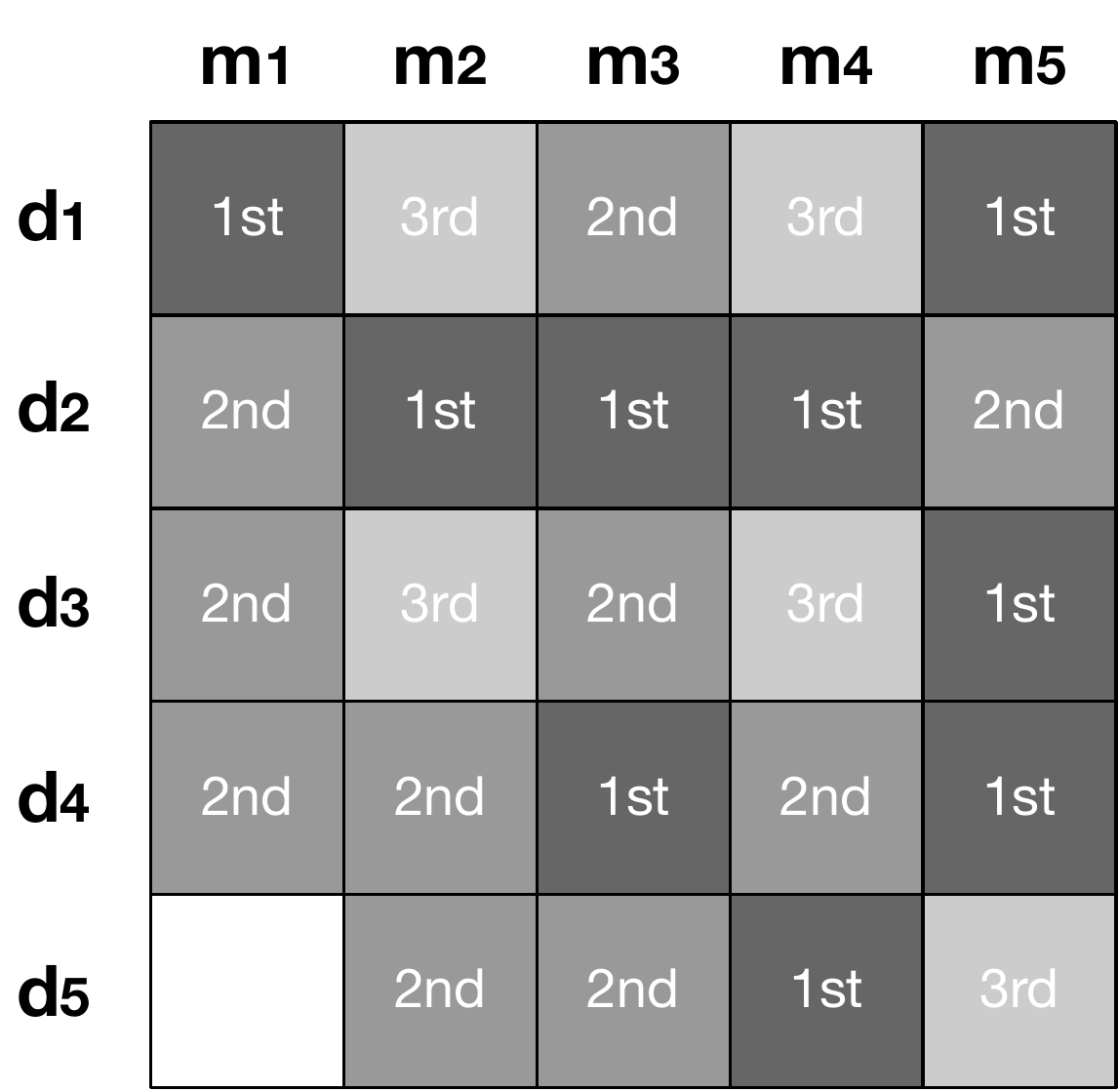}
        \caption{}
        \label{fig:high_order_matrix}
    \end{subfigure}
    \hfill
    \caption{Illustration of the $l$-order proximity matrix via random walk. Figure~\ref{fig:first_order_matrix} shows a 1-order matrix and Figure~\ref{fig:high_order_matrix} shows a 3-order proximity matrix for the PHA installation graph shown in Figure~\ref{fig:motivation}. We use grayscale to illustrate the diminishing correlation among vertices (see Figure~\ref{fig:neighbour_of_neighbour}). A blank cell implies that $l$-order proximity is 0. As it can be seen, the 3-order proximity matrix (Figure~\ref{fig:high_order_matrix}) offers more information comparing to 1-order proximity matrix (Figure~\ref{fig:first_order_matrix}).}
    \label{fig:proximity_matrix}
\end{figure}

The key challenge of \approach is building a model that learns both explicit relations between devices and PHAs, and implicit relationships among PHAs from the global PHA installation graph, and predict the future installations. 
To address this challenge, we use a graph representation learning model~\cite{yang2018hop}. 
Representation learning~\cite{bengio2013representation,yang2018hop} is able to extract useful information to build prediction models \emph{without feature engineering}. 
This characteristic of representation learning is particularly desirable since it enables us to understand the explicit and implicit dynamics between devices and PHAs without a time-consuming feature engineering step.   

Random walk has been proven to provide insightful transitive associations between two vertices in a graph, and has been successfully applied in numerous settings, \eg community detection~\cite{pons2005computing,rosvall2008maps}, personalized PageRank~\cite{page1999pagerank}, and Sybil account detection~\cite{danezis2009sybilinfer,boshmaf2015integro,stringhini2015evilcohort}. The core of \approach is building a high-order proximity matrix by conducting truncated random walks with a decay factor on the global PHA installation graph. This proximity matrix captures both direct and indirect relationships between devices and PHAs, and, at the same time, discriminates the strength between different orders of proximity (see Section~\ref{sec:motivation}) due to the decay factor. The low dimensional representations are later learned by factorizing this high-order proximity matrix using random walk approximation techniques. We provide the following definition to describe the technical details of \approach in formal terms.

\noindent\textbf{$l$-order proximity.} Let $\mathbf{RW_{\mathbf{d}_0}} = (\mathbf{d}^0, \mathbf{m}_i^1, \mathbf{d}_j^1, \mathbf{m}_z^2,$ $ ..., \mathbf{d}_h^{l-1}, \mathbf{m}_k^l, ...)$ denote a random walk starting from a device vertex $\mathbf{d}_0$ on a PHA installation graph $\mathbf{G}=(\mathbf{D} \cup \mathbf{M}, \mathbf{E})$, where superscript denotes the rank of vertex occurrence in the random walk. $l$-order proximity between $\mathbf{d}_0 \in \mathbf{D}$ and $\mathbf{m}_k^l \in \mathbf{M}$ is defined by a decay function $C$ (see Eq~\ref{eq:obj_func}) of the rank of occurrence (\ie $C(l)$) and the transition probability between them in the random walk. If there is no walkable path from $\mathbf{d}$ to $\mathbf{m}_k^l$ , the $l$-order proximity is 0.
     
\noindent \textbf{Example.} Given a random walk $RW_{\mathbf{d}_{1}^{0}} = (\mathbf{d}_{1}^{0}, \mathbf{m}_{5}^{1}, \mathbf{d}_{4}^{1}, \mathbf{m}_{3}^{2})$ as illustrated in \ding{192} Figure~\ref{fig:rw_prediction}, $\mathbf{m}_{3}$ is at 2-order proximity of $\mathbf{d}_1$. Similarly, given a random walk $RW_{\mathbf{d}_{3}^{0}} = (\mathbf{d}_{3}^{0}, \mathbf{m}_{5}^{1}, \mathbf{d}_{4}^{1}, \mathbf{m}_{3}^{2}, \mathbf{d}_{2}^{3}, \mathbf{m}_{2}^{3})$ as illustrated in \ding{193} Figure~\ref{fig:rw_prediction},  $\mathbf{m}_{2}$ is at 3-order proximity of $\mathbf{d}_3$.

Figure~\ref{fig:proximity_matrix} shows two $l$-order proximity matrices induced by random walks. Figure~\ref{fig:first_order_matrix} shows an example of 1-order proximity matrix and Figure~\ref{fig:high_order_matrix} shows a 3-order proximity matrix induced by random walks on the PHA installation graph (see Figure~\ref{fig:motivation}). We can see that, with the increasing value of $l$, $l$-order proximity captures more high-order relationship information between devices and PHAs.

Given the $l$-order proximity matrix, formally, the objective function for learning low dimensional representations (\ie $\Phi$) is defined in Eq~\ref{eq:obj_func}

\begin{eqnarray} 
\label{eq:obj_func}
{\scriptstyle \mathcal{L}} &  = &   {\scriptstyle \sum_{\substack{1 \leq l \leq K \\ d_i, (m_j, m_{j'})}} C(l) \mathbb{E}_{\substack{m_j \sim P_{d_i}^l \\ m_{j'} \sim P_N}} \left[ \mathcal{F}(\phi_{d_i}^T \phi_{m_{j'}}, \phi_{d_i}^T \phi_{m_{j}}) \right]}  \nonumber \\
 & &  {\scriptstyle + \lambda_{\Phi} \parallel \Phi \parallel_{2}^2} 
\end{eqnarray}

\noindent where $C(l)=1/l$ denotes the decay function, $P_{d_i}^l(\cdot)$ denotes the $l$-order probability distribution of a PHA $m_j$ sampled from a random walk $\mathbf{RW_{\mathbf{d}_i}}$ (see Eq~\ref{eq:rw_prob}), $P_N$ denotes a uniform distribution of all items from which a PHA $i'$ was drawn,  and $\mathcal{F}(\phi_{d_i}^T \phi_{m_{j'}}, \phi_{d_i}^T \phi_{m_{j}})$ is a ranking objective function discriminating the observed PHAs installations (\ie $\phi_{d_i}^T \phi_{m_{j}}$) from unobserved PHA installations (\ie $\phi_{d_i}^T \phi_{m_{j'}}$) in the low dimensional embedding space (see Eq~\ref{eq:rank_func}). Eq~\ref{eq:rw_prob} is formalized as:

\begin{equation}
\label{eq:rw_prob}
P_{v_x}^l(v_y)=\left\{
\begin{array}{c l}    
     \frac{\mathbf{A}_{v_x,v_y} deg(v_y)}{\sum_{v_{y'}} \mathbf{A}_{v_x,v_{y'}} deg(v_{y'})} & l=1, v_x \in \mathbf{D}\\
     \frac{\mathbf{A}_{v_y,v_x} deg(v_y)}{\sum_{v_{y'}} \mathbf{A}_{v_{y'},v_x} deg(v_{y'})} & l=1, v_x \in \mathbf{M}\\
     p_{v_x}^1(v_\alpha) p_{v_\alpha}^{l-1}(v_\beta) p_{v_\beta}^1(v_y) & otherwise
\end{array}\right.
\end{equation}

\noindent approximates the probability of sampling a $l$-th neighbor $v_y$ for $v_x$ given a random walk $(v_x, v_\alpha, ..., v_\beta, v_y)$, where $v_\alpha$ denotes the vertex after visiting $v_x$ and $v_\beta$ denotes the vertex before $v_y$. In this way, it simplifies the cumulative process of counting all intermediate vertices from all possible random walks from $v_x$ to $v_y$ in a recursive manner. Note that if $v_x \in \mathbf{D}$, then $v_\alpha \in \mathbf{M}$. Otherwise, if $v_x \in \mathbf{M}$, then $v_\alpha \in \mathbf{D}$. For example, given a random walk $RW_{\mathbf{d}_{1}^{0}} = (\mathbf{d}_{1}^{0}, \mathbf{m}_{5}^{1}, \mathbf{d}_{4}^{1}, \mathbf{m}_{3}^{2})$ as illustrated in \ding{192} Figure~\ref{fig:rw_prediction}, $v_x$ is $\mathbf{d}_{1}^{0}$,  $v_\alpha$ is $\mathbf{m}_{5}$, $v_\beta$ is $\mathbf{d}_{4}$ and $v_y$ is $\mathbf{m}_{3}^{2})$.

\noindent We define the ranking objective function $\mathcal{F}(\cdot)$ as follows:

\begin{equation}
\label{eq:rank_func}
     \mathcal{F}(\phi_{d_i}^T \phi_{m_{j'}}, \phi_{d_i}^T \phi_{m_{j}}) = \mathbbm{1}_{(\delta > \frac{\epsilon}{k})} log(\delta)
\end{equation}

\noindent where $\delta = \phi_{d_i}^T \phi_{m_{j'}} - \phi_{d_i}^T \phi_{m_{j}}$ and $\mathbbm{1}_{\zeta}: \zeta \rightarrow \{ 0, 1 \}$ denotes an indicator function for condition $\zeta$, where $\zeta=\delta > \frac{\epsilon}{k}$. we employ a random walk approximation technique~\cite{chen2017vertex} to approximate the matrix factorization results. Eq~\ref{eq:obj_func} is accordingly minimized using asynchronous stochastic gradient descent~\cite{recht2011hogwild} in the paper.

\section{Dataset}
\label{sec:dataset}

\begin{table}[t]
\centering
\resizebox{0.9\linewidth}{!} {
\begin{tabular}{|c|c|c|c|c|c|}
\hline
\textbf{Dataset} & \textbf{Date(s)} & \textbf{\# Dev.} & \textbf{\# PHA} & \textbf{V}  & \textbf{E}  \\ \hline
\textbf{$DS_1$} & Mar 1 2019 & 644,823 & 63,650 & 708,473 & 844,531  \\ \hline
\textbf{$DS_2$} & Mar 1-14 2019 & 1,272,505 & 99,464 & 1,371,969 & 1,675,952   \\ \hline
\textbf{$DS_3$} & Mar 1-31 2019 & 1,864,021 & 131,903 & 1,995,924 & 2,509,744   \\ \hline
\end{tabular}
}
\caption{Summary of experimental datasets.}
\label{tab:exp_datasets}
\end{table}

To evaluate \approach's performance, we focus on Android PHA detection data collected by a major security company's mobile security product.
This company offers end users to explicitly opt in to its data sharing program during the setup process when the app is run for the first time to help improving its detection capabilities. 
The end users are shown a dialog about the purpose of the telemetry collection in the license agreement, and how the global privacy policy of the security company safeguards the data.
The license agreement specifies that the telemetry is ``processed for the purposes of delivering the product by alerting the User to potentially malicious applications as well as improving the app security feature'' and is ``kept in an encrypted pseudonymized form.''
To preserve the anonymity of users and their devices, client identifiers are anonymized and it is not possible to link the collected data back to the users and the mobile devices that originated it. 
The mobile security app only collects detection metadata (\eg anonymized device identifier, timestamp, file SHA2, etc.), and it cannot inspect network traffic data, hence the company does not collect any actual communication/user data, or other types of PII. 
The telemetry used in this paper does not contains any personally identifiable information (PII).
The product runs periodically scanning the apps that the user installed and records meta information associated with PHAs that are detected based on pre-defined signatures or whose behavior violates the pre-defined policies. 
From this dataset, we extract the following information: anonymized device identifier, timestamp, file SHA2, PHA package name, and other detection information (\eg installation types, installer packages).   

We collect 31 days of data over March 2019 to thoroughly investigate the prediction effectiveness, stability, and resilience of a real world deployment of \approach. 
On average, we collect 2.4 million raw detection events from 1.8 million end users (the majority of the users run Android 6.0 Marshmallow or above) located in over 200 countries and territories per day. 
From these 31 days of raw data, we build three datasets summarized in Table~\ref{tab:exp_datasets}. 
$DS_1$ is a single day data using raw data collected on March 1 2019. 
$DS_2$ and $DS_3$ are one week data (between March 1 and March 7 2019) and one month data (between March 1 and March 31 2019). 
We use all three datasets to compare \approach's performance to those of the baseline methods (see Section~\ref{sec:exp_comparison}).
We later focus on $DS_2$ and $DS_3$ and use them to evaluate \approach's prediction and runtime performance given different real world scenarios.  

\noindent \textbf{Reproducibility.} 
Note that \approach can operate on fully anonymized data (\eg device identifiers and PHA packages names can be replaced by other unique identification schemes) to protect the end users' privacy. 
It is straightforward for the researchers to reproduce \approach's results by generating bipartite networks matching the characteristics stated in Table~\ref{tab:exp_datasets} and Figure~\ref{fig:malware_degree_distribution}.

\begin{table*}[h]
\centering
\resizebox{0.9\linewidth}{!} {
\begin{tabular}{|c|c|c|c|c|c|c|c|c|c|c|}
\hline
 & \multicolumn{5}{c|}{\textbf{Training}} & \multicolumn{5}{c|}{\textbf{Test}} \\ \hline
\textbf{Dataset} & \textbf{Period} & \textbf{Ratio} & \textbf{\# Events} & \textbf{\# Dev} & \textbf{\# Apps} & \textbf{Period} & \textbf{Ratio} & \textbf{\# Events} & \textbf{\# Devs} & \textbf{\# Apps} \\ \hline
$DS_1$ & 00:00 - 18:00 (Mar. 1) & 0.73 & 844,531 & 644,823 & 63,650 & 18:00 - 24:00 (Mar. 1) & 0.27 & 317,474 & 189,327 & 26,083 \\ \hline
$DS_2$ & March 1 - 6 & 0.86 & 2,050,865 & 1,272,505 & 99,464 & March 7 & 0.14 & 334,383 & 237,594 & 32,961 \\ \hline
$DS_3$ & March 1 - 24 & 0.84 & 3,194,838 & 1,864,021 & 131,903 & March 25 - 31 & 0.16 & 599,458 & 404,417 & 47,099 \\ \hline
\end{tabular}
}
\caption{Training/test data split. We deliberately eliminate the temporal bias~\cite{pendlebury2019tesseract} that may be introduced using traditional training/test split.}
\label{fig:training_test_split}
\end{table*}

\noindent \textbf{Data Limitations.} It is important to note that the PHA detection data is collected passively. That is, a PHA detection event is recorded when the product detects a potentially harmful application that matches a pre-defined signature including its behavior, communications, and policy violations. Any events pre-emptively blocked by other security products (\eg application store link blacklists, cloud-based app reputation system) cannot be observed. Additionally, any PHAs that did not match the predefined signatures are also not observed. The prediction model used in this paper can, therefore, predict the PHA observed by the mobile security product from this security company. We discuss more details on the limitations underlying the data used by \approach in Section~\ref{sec:discussion}.

\section{Evaluation}
\label{sec:evaluation}

In this section, we describe the experiments used to evaluate \approach. We designed several experiments that allow us to answer the following research questions:

\noindent \textbf{RQ1:} What is \approach's performance in predicting the upcoming PHA installation events and how does its performance compare to baseline and state-of-the-art methods? \newline
\noindent \textbf{RQ2:} Can we interpret and explain \approach's prediction performance?\newline
\noindent \textbf{RQ3:} How does \approach cope with data latency issues that arise in real world deployment? In other words, does \approach still work if mobile devices report data about the infections that they encountered with a significant delay?\newline
\noindent \textbf{RQ4:} What is \approach's prediction performance when we need to retrain the model periodically?\newline
\noindent \textbf{RQ5:} Is \approach's runtime performance acceptable for a real world deployment?

\subsection{Experimental Setup}
\label{sec:exp_setup}

\noindent \textbf{Implementation.} We implement \approach's graph representation learning component in C++ using the OpenMP~\cite{dagum1998openmp} library, the PHA graph reconstruction component using Python 3.7.2 and the NetworkX library, and employ the Random Forest~\cite{breiman2001random} implementation from Sklearn as the prediction engine. In terms of baseline implementations (see Section~\ref{sec:exp_comparison}), preferential attachment is implemented in Python 3.7.2 and the rest are implemented in C++. All experiments are carried out on a server with dual 2.67GHz Xeon CPU X5650 and 256GB of memory.

\noindent \textbf{Training/test data split.} For $DS_1$, all methods are trained on the first 18 hours data (00:00 - 18:00) and tested on the following 6 hours data (18:00 - 24:00). For $DS_2$, all methods are trained on the first 6 days of data (March 1 - March 6 2019) and tested on the 7th day of data (\ie March 7 2019). For $DS_3$, all methods are trained on the first 3 weeks of data (March 1 - March 24 2019) and tested on the 4th week of data (\ie March 25 - 31 2019). In this setup, we deliberately eliminate the temporal bias that may be introduced using traditional training/test splits~\cite{pendlebury2019tesseract}. However, different training and test ratios are therefore introduced due to the fact that PHA installations are not evenly distributed. Details of the training and test data are summarized in Table~\ref{fig:training_test_split}. 

\noindent \textbf{Prediction formulation.} We formulate our prediction task as a binary prediction problem for edges in the network where two classes are considered: positive (\ie observed edges) and negative (\ie absence of edges). For both training and test data, we sample an equal number of non-existing edges~\cite{pendlebury2019tesseract} from the observed PHA installation edges to form the negative class. In this way, we are able to generalize \approach's performance by evaluating its performance on a set of data not used for training, which is assumed to approximate the typical unseen data that the model will encounter in the wild, reducing the chance of overfitting.

\noindent \textbf{Choosing the prediction model.} A plethora of prediction models for supervised learning exist. Usually, some work better than others for a specific dataset or domain. In this paper, we experimented with several different prediction models, and select Random Forest~\cite{breiman2001random} as the prediction engine due to its scalability, bias-variance trade-off, and overall good performance given all methods. We use 10-fold cross validation for all methods. We also provide a detailed discussion in Section~\ref{sec:discussion} and examine how different models may impact the prediction results. 

\noindent \textbf{Hyperparameters.} We use grid search on $DS_1$ to identify the best parameters. Experimentally, we set the embedding size $d$ to 128, the high-order proximity threshold $l$ to 4 (effectively, the random walk was truncated at length 6), and negative sampling times~\cite{chen2017vertex} to 50. For all baselines, we use the best parameters stated in the respective papers. We also set the number of trees in the random forest algorithm to 20. Note that hyperparameters optimization remains an active research area in the machine learning community. We refer the reader to~\cite{bergstra2011algorithms} for an overview.

\noindent \textbf{Evaluation metrics.} 
We use the true positive rate (TPR)~\cite{provost1998glossary}, the false positive rate (FPR)~\cite{provost1998glossary}, the area under the ROC curve (AUC) score~\cite{hanley1982meaning}, and the average precision (AP)~\cite{kishida2005property,manning2008introduction} to extensively evaluate \approach's performance.
TP, TN, FP, and FN in the rest of the section denote the number of true positives, true negatives, false positives, and false negatives, respectively.

\noindent \textbf{Thresholds.} Note that predicting the exact PHA installations for end users is a challenging task in the real world. A higher false positive rate leads to worse user experience hence higher customer churn rate. For example, a 0.01 FPR for 317,474 events (see $DS_1$, Table~\ref{fig:training_test_split}) may have an undesirable impact on approximately 3,174 end users in a six hour window which is not acceptable in a real world deployment. We therefore fix the FPR at three scales, respectively 0.0001, 0.001 and 0.005 from real world usability perspective.

\begin{table*}[ht]
\centering
\resizebox{0.9\linewidth}{!} {
\begin{tabular}{|c|c|c|c|c|c|c|c|c|c|c|c|c|c|c|c|}
\hline
 & \multicolumn{5}{c|}{$DS_1$} & \multicolumn{5}{c|}{$DS_2$} & \multicolumn{5}{c|}{$DS_3$} \\ \hline
\textbf{Method} & \textbf{\begin{tabular}[c]{@{}c@{}}TPR\\ @\\ 0.0001\end{tabular}} & \textbf{\begin{tabular}[c]{@{}c@{}}TPR\\ @\\ 0.001\end{tabular}} & \textbf{\begin{tabular}[c]{@{}c@{}}TPR\\ @\\ 0.005\end{tabular}} & \textbf{\begin{tabular}[c]{@{}c@{}}ROC \\ AUC\end{tabular}} & \textbf{\begin{tabular}[c]{@{}c@{}}AP\end{tabular}} & \textbf{\begin{tabular}[c]{@{}c@{}}TPR\\ @\\ 0.0001\end{tabular}} & \textbf{\begin{tabular}[c]{@{}c@{}}TPR\\ @\\ 0.001\end{tabular}} & \textbf{\begin{tabular}[c]{@{}c@{}}TPR\\ @\\ 0.005\end{tabular}} & \textbf{\begin{tabular}[c]{@{}c@{}}ROC\\ AUC\end{tabular}} & \textbf{\begin{tabular}[c]{@{}c@{}}AP\end{tabular}} & \textbf{\begin{tabular}[c]{@{}c@{}}TPR\\ @\\ 0.0001\end{tabular}} & \textbf{\begin{tabular}[c]{@{}c@{}}TPR\\ @\\ 0.001\end{tabular}} & \textbf{\begin{tabular}[c]{@{}c@{}}TPR\\ @\\ 0.005\end{tabular}} & \textbf{\begin{tabular}[c]{@{}c@{}}ROC\\ AUC\end{tabular}} & \textbf{\begin{tabular}[c]{@{}c@{}} AP \end{tabular}} \\ \hline
\textbf{Pref. Attach.} & 0.072 & 0.268 & 0.512 & 0.977 & 0.974 & 0.099 & 0.310 & 0.593 & 0.980 & 0.978  & 0.094 & 0.338 & 0.584 & 0.981 & 0.980 \\ \hline
\textbf{1st-order prox.} & 0.782 & 0.898 & 0.936 & 0.983 & 0.986 & 0.837 & 0.927 & 0.950 & 0.982 & 0.986 & 0.844 & 0.927 & 0.965 & 0.990 & 0.990 \\ \hline
\textbf{2nd-order prox.} & 0.863 & 0.922 & 0.959 & 0.992 & 0.993 & 0.867 & 0.918 & 0.953 & 0.993 & 0.993 & 0.868 & 0.941 & 0.966 & 0.993 & 0.994 \\ \hline
\textbf{high-order prox.} & 0.873 & 0.969 & 0.985 & 0.997 & 0.997 & 0.893 & 0.957 & 0.977 & 0.996 & 0.996 & 0.879 & 0.951 & 0.978 & 0.995 & 0.996 \\ \hline
\textbf{\approach} & \textbf{0.991} & \textbf{0.996} & \textbf{0.998} & \textbf{0.999} & \textbf{0.999} &
 \textbf{0.994} & \textbf{0.997} & \textbf{0.998} & \textbf{0.999} & \textbf{0.999} 
 & \textbf{0.992} & \textbf{0.996} & \textbf{0.997} & \textbf{0.999} & \textbf{0.999} \\ \hline
\end{tabular}
}
\caption{Prediction performance comparison study: \approach vs. baseline approaches. }
\label{tab:exp_comparison}
\end{table*}

\begin{figure*}[t]
     \centering
	 \resizebox{0.9\linewidth}{!} {
    \begin{subfigure}[t]{0.33\textwidth}
        \includegraphics[width=\textwidth]{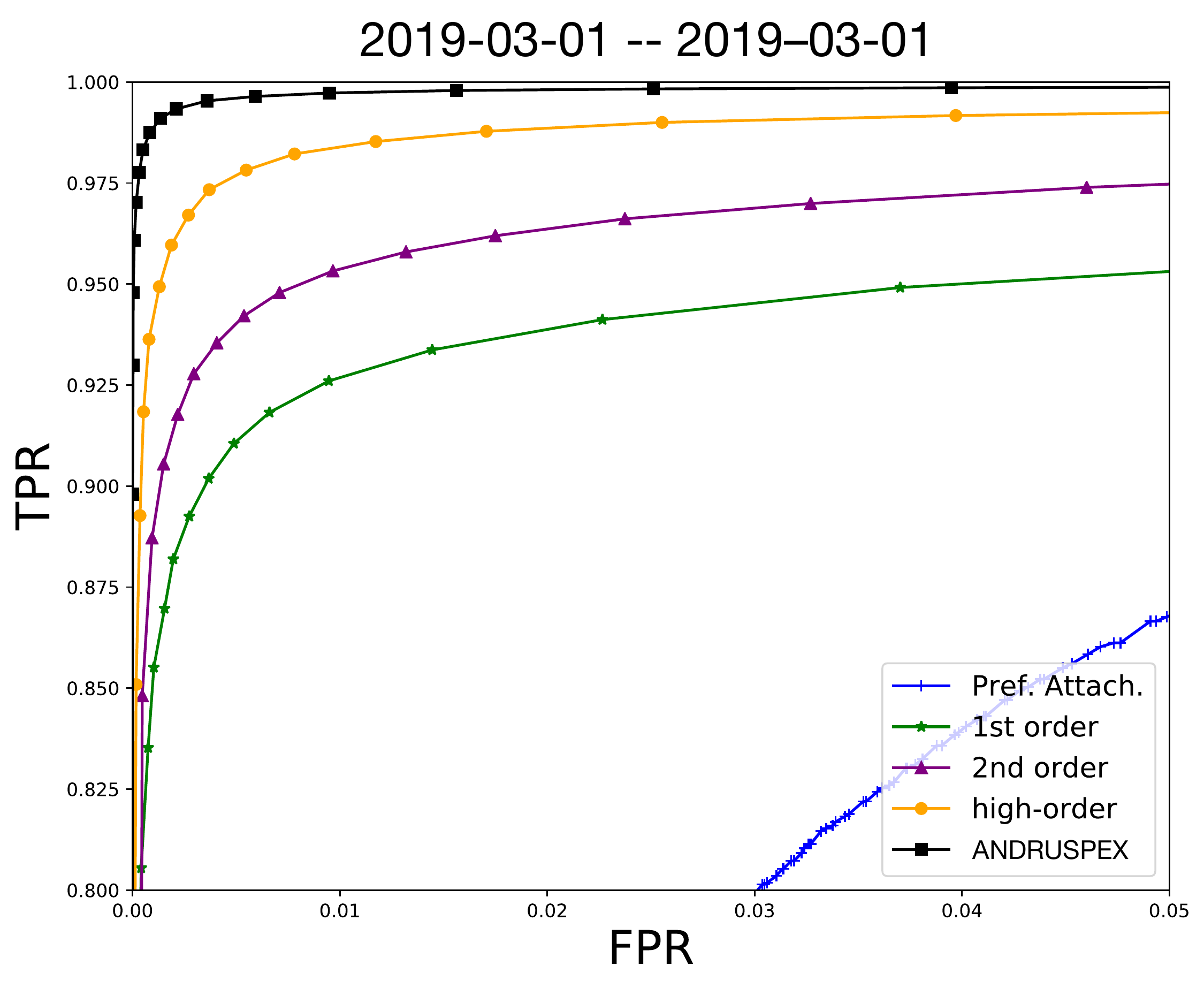}
        \caption{$DS_1$}
    \end{subfigure}
    \hfill
    \begin{subfigure}[t]{0.33\textwidth}
        \includegraphics[width=\textwidth]{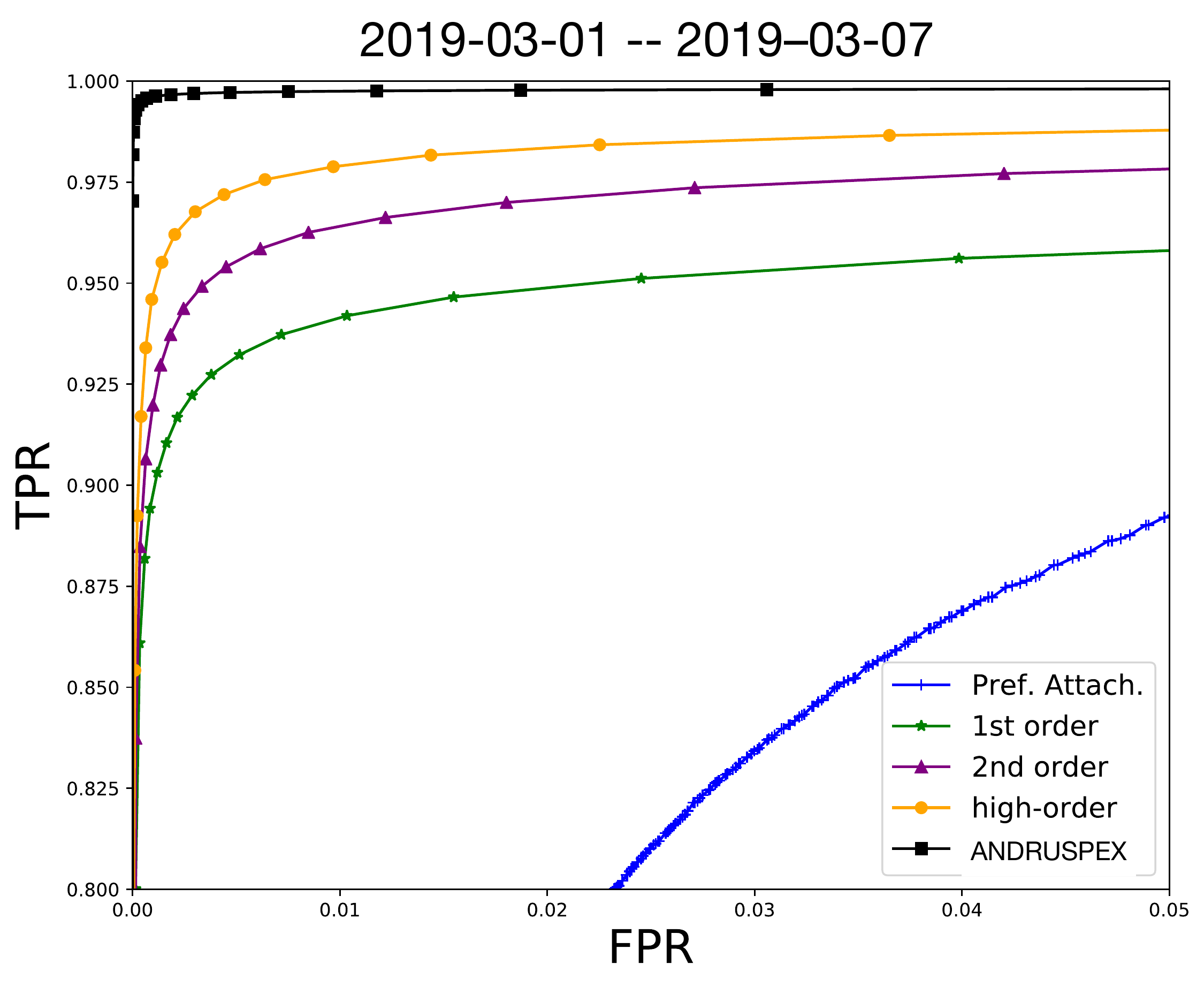}
        \caption{$DS_2$}
    \end{subfigure}
    \hfill
    \begin{subfigure}[t]{0.33\textwidth}
        \includegraphics[width=\textwidth]{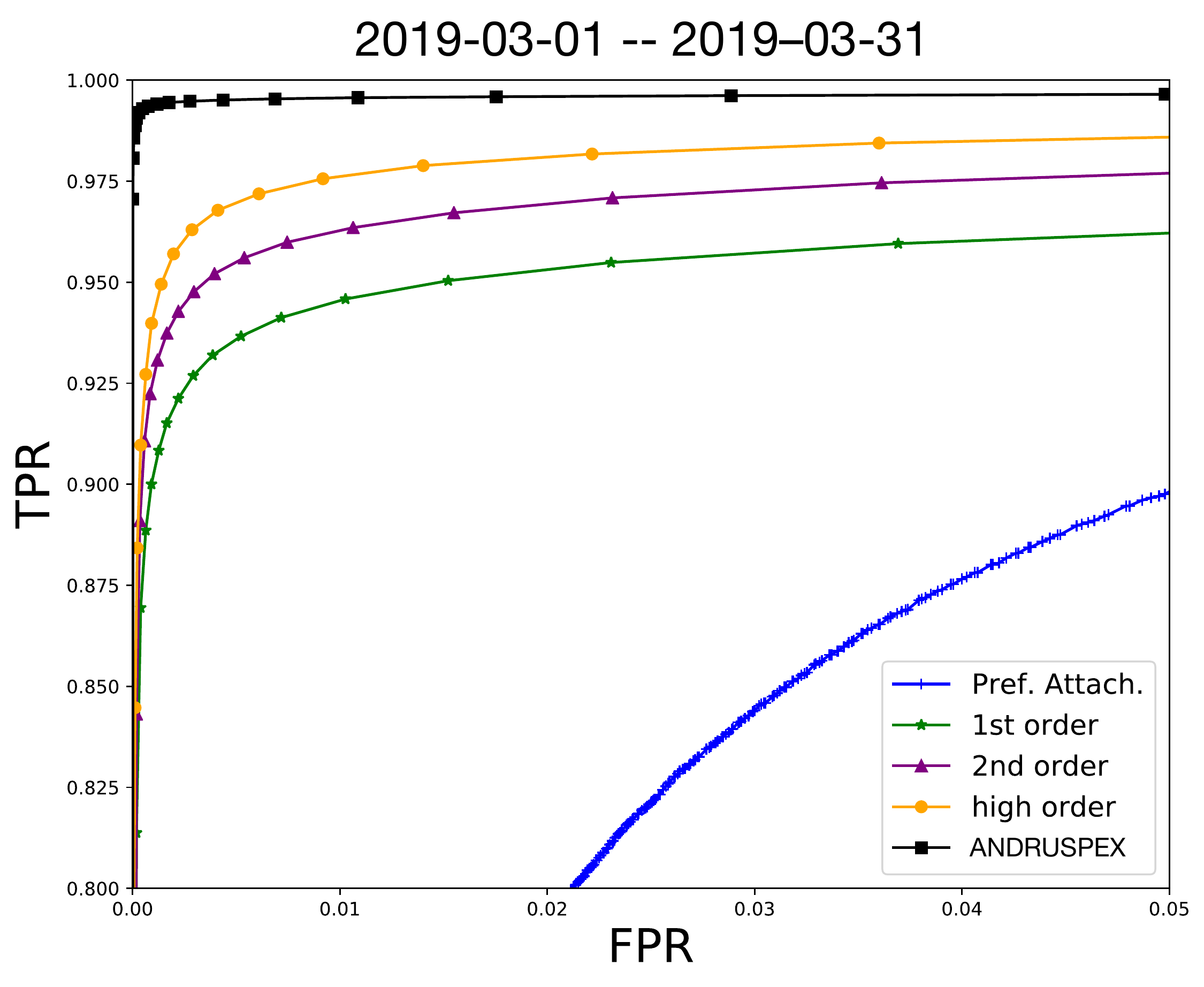}
        \caption{$DS_3$}
    \end{subfigure}
	}
    \hfill
	\caption{ROC curves comparison: \approach vs. baseline approaches.}
	\label{fig:exp_comparison_auc}
\end{figure*}

\begin{figure*}[ht]
     \centering
    \begin{subfigure}[t]{0.3\textwidth}
        \includegraphics[width=\textwidth]{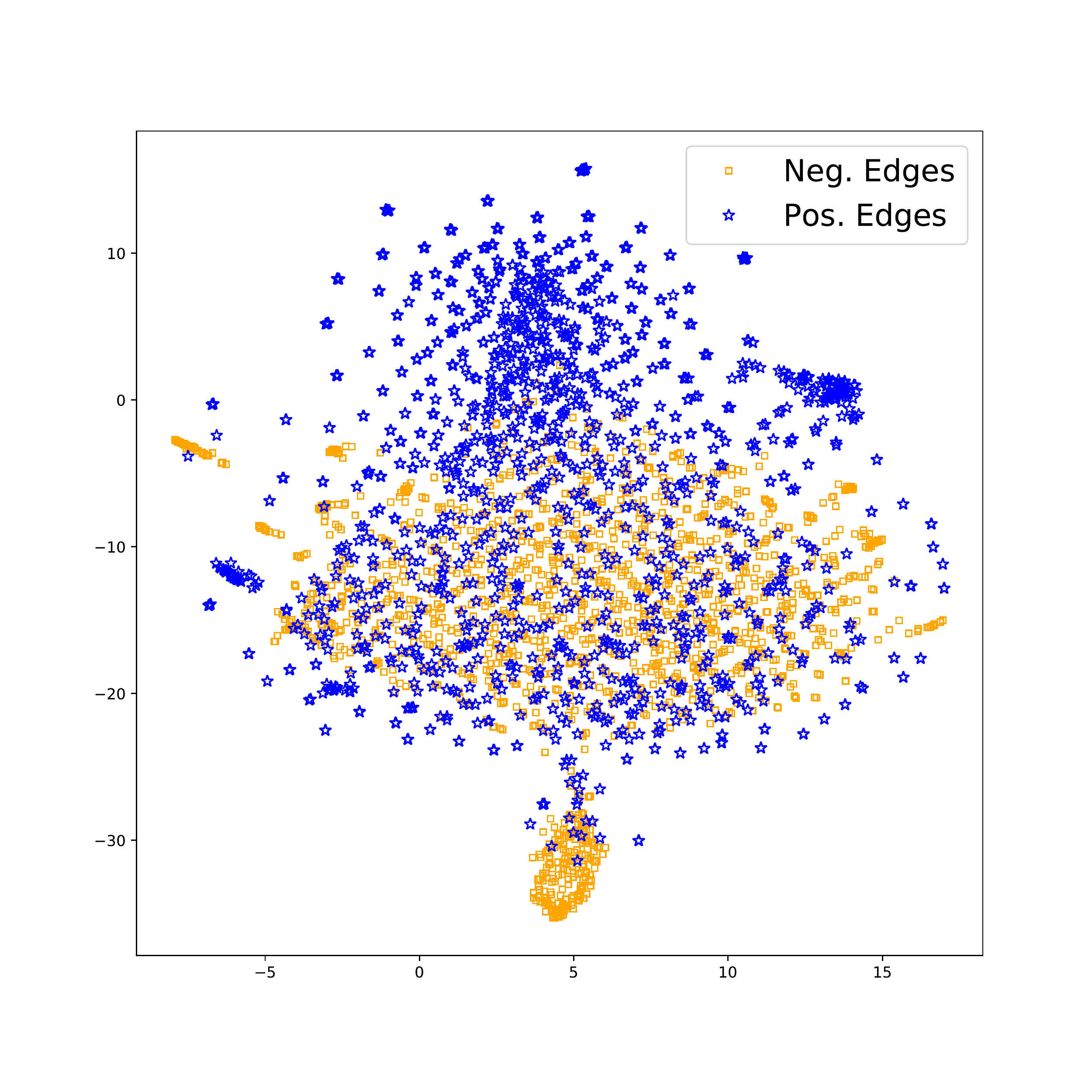}
        \caption{1st-order proximity}
        \label{fig:largevis_lineo1}
    \end{subfigure}
    \hfill
    \begin{subfigure}[t]{0.3\textwidth}
        \includegraphics[width=\textwidth]{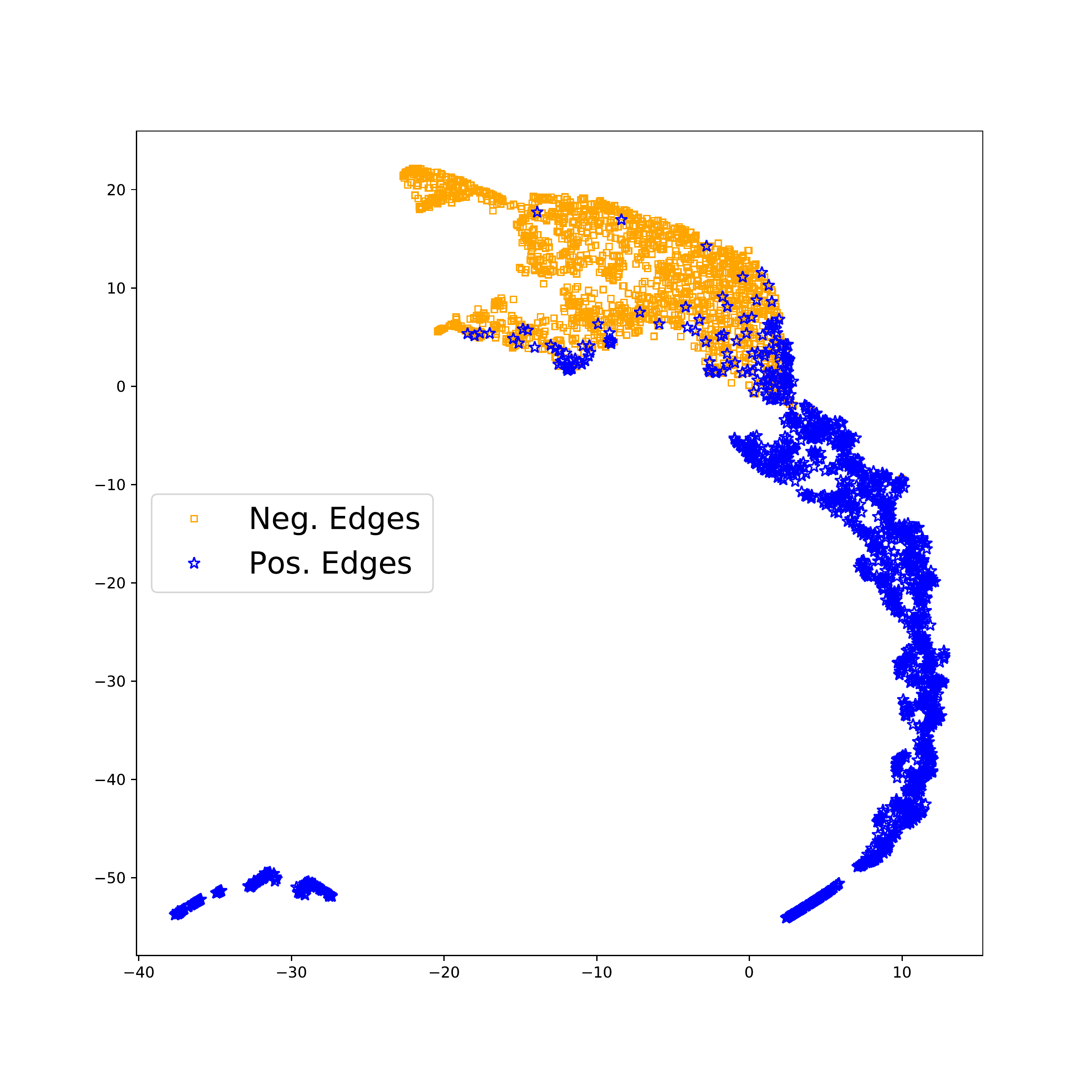}
        \caption{2nd-order proximity}
        \label{fig:largevis_lineo2}
    \end{subfigure}
    \hfill
    \begin{subfigure}[t]{0.3\textwidth}
        \includegraphics[width=\textwidth]{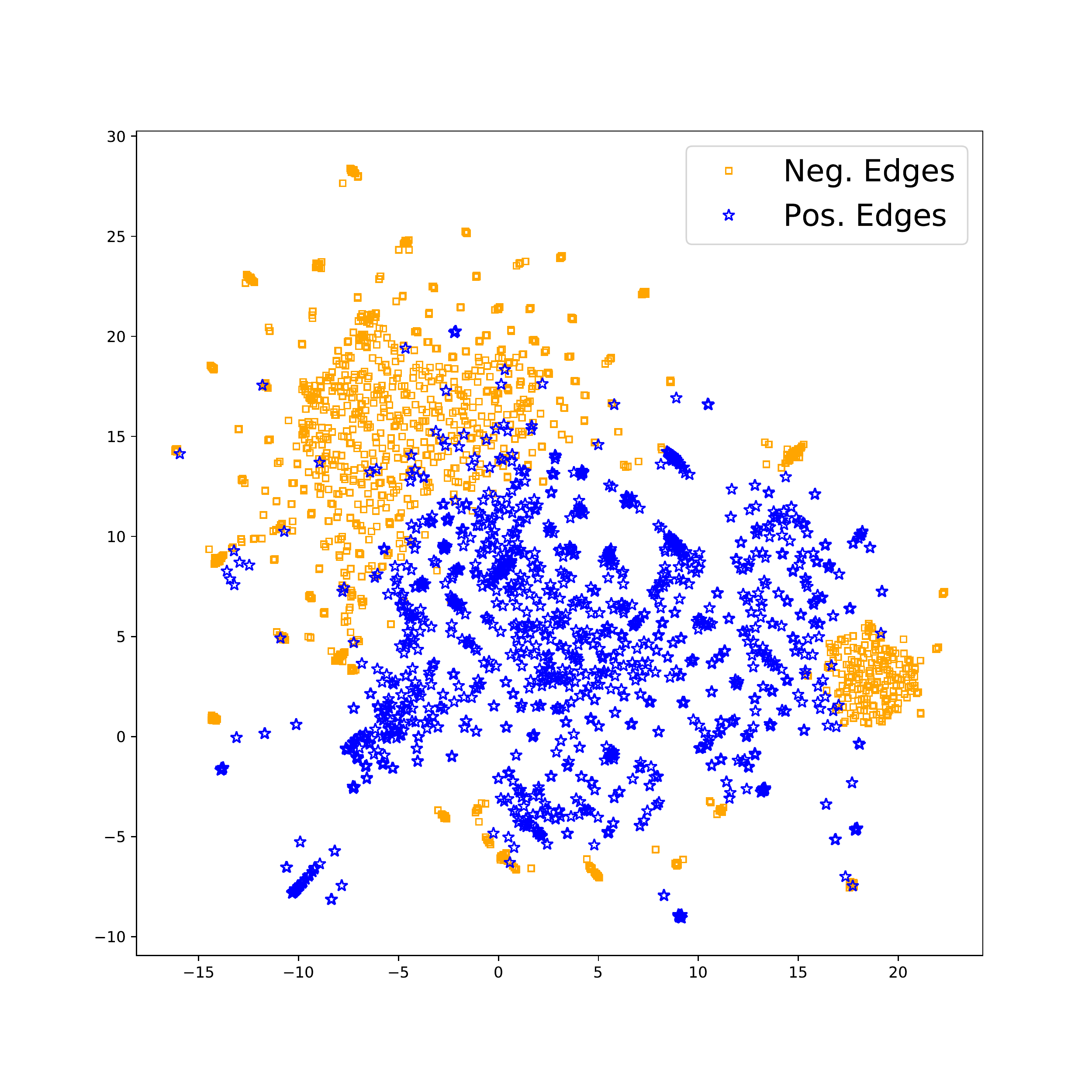}
    \caption{high-order proximity} 
    \label{fig:lagevis_cse}
    \end{subfigure}
    \hfill
    \begin{subfigure}[t]{0.3\textwidth}
        \includegraphics[width=\textwidth]{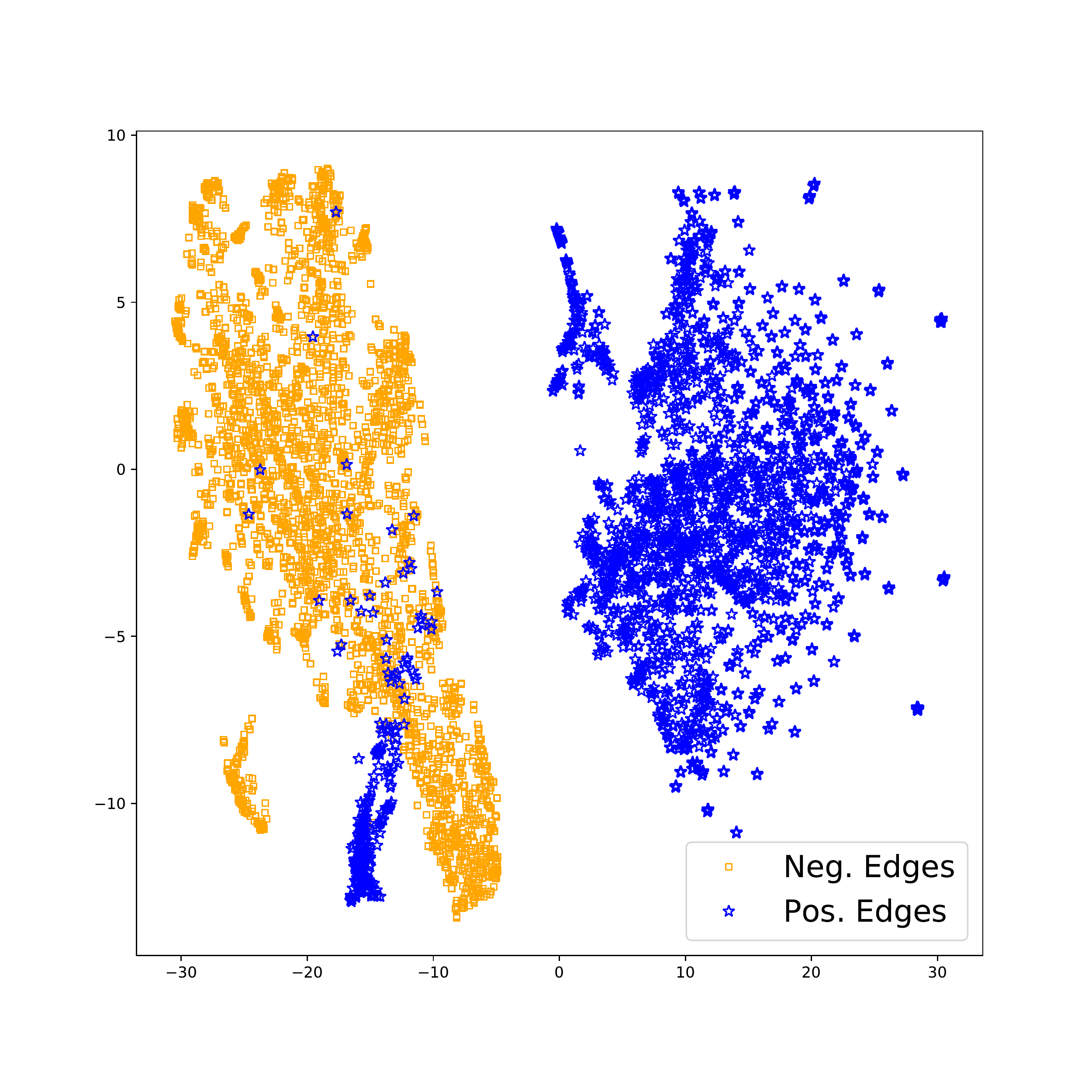}
    \caption{\approach (using $DS_1$)} 
    \label{fig:lagevis_approach}
    \end{subfigure}
    \hfill
    \begin{subfigure}[t]{0.3\textwidth}
        \includegraphics[width=\textwidth]{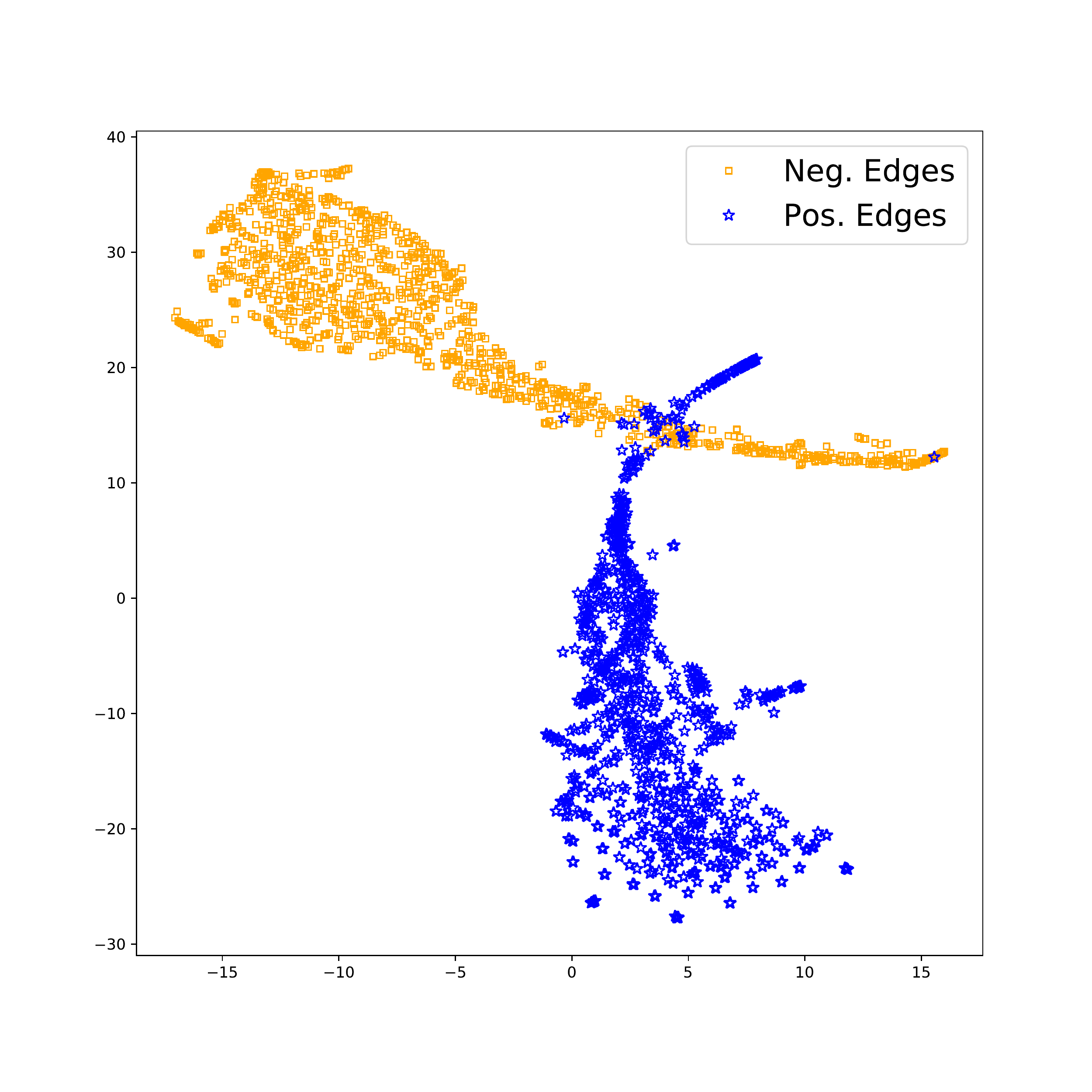}
    \caption{\approach (using $DS_2$)} 
    \label{fig:lagevis_approach_1w}
    \end{subfigure}
    \hfill
    \begin{subfigure}[t]{0.3\textwidth}
        \includegraphics[width=\textwidth]{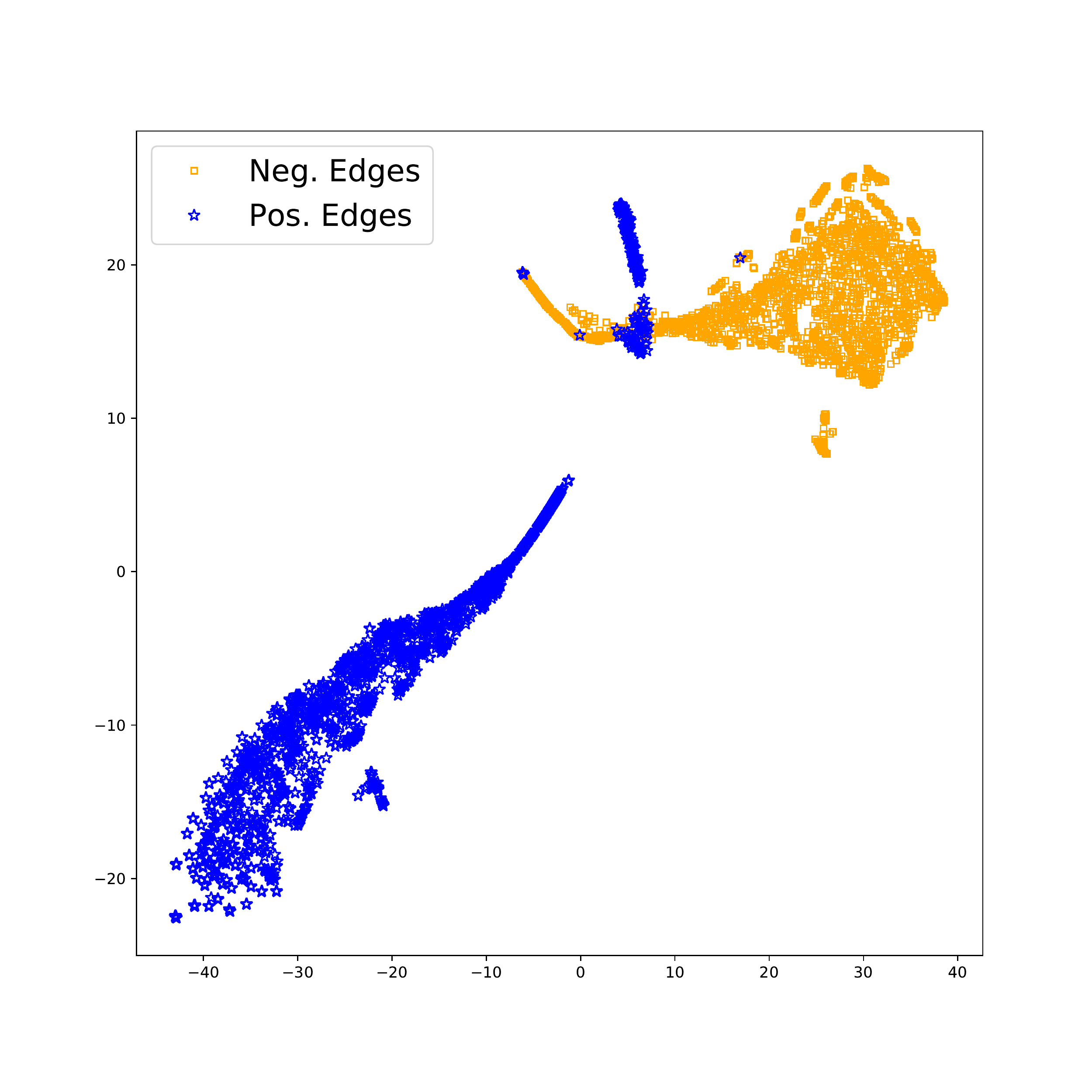}
    \caption{\approach (using $DS_3$)} 
    \label{fig:lagevis_approach_1m}
    \end{subfigure}
    \hfill
    \caption{LargeVis projections of positive (\ie observed) and negative (\ie non-existent) PHA installation events (\ie concatenated $2d$-dimensional vectors) from all methods (see Section~\ref{sec:exp_comparison}). The visualization results demonstrate that \approach can offer a good PHA installation event representation in the latent space.  Note that preferential attachment does not produce embeddings hence is excluded from the visualization. This figure is better viewed in color.}
    \label{fig:exp_largevis}
\end{figure*}

\subsection{Comparison Study (RQ1)}
\label{sec:exp_comparison}

\noindent \textbf{Experimental setup.} We use all three datasets (see Table~\ref{tab:exp_datasets}) in this section to comparatively study prediction performance of baseline methods and \approach. Training and test data are split following the setup specified in Section~\ref{sec:exp_setup}.

\noindent \textbf{Baselines.} In this section we aim at understanding whether \approach is required for the task of predicting PHA installations, or whether existing baseline methods are sufficient for the task at hand. For comparison purposes, we implemented the following four baseline methods.

\noindent \emph{Preferential Attachment}. Preferential attachment~\cite{barabasi2002evolution,newman2001clustering} is a popular network growth model, in which vertices accumulate new edges in proportion to the number of the edges they already have. 
The rationale of using preferential attachment as a baseline is to verify if popular PHAs dominate the future PHA installations.  
In this paper, we use PA\_score$(v_i, v_j) = \frac{|\mathcal{N}(v_i)| \times |\mathcal{N}(v_j)|}{2 \times |\mathcal{E}|}$ to compute the normalized preferential attachment score of all vertex pairs.

\noindent \emph{First-order and Second-order proximity}. We use LINE~\cite{tang2015line} as the first-order and second-order proximity baseline. This method directly generates vertex embeddings by taking explicit relationships (first-order proximity) and indirect relationships (second-order proximity) into consideration. 
It is able to partially preserve the local and global structure of the graph by conducting edge sampling. 
The rationale of using LINE is to verify if a limited number of proximity can already offer good prediction results. 

\noindent \emph{High-order proximity via matrix factorization}. Chen~\etal~\cite{chen2019collaborative} provided a collaborative similarity embedding (CSE) framework to model high-order proximity by capturing the similarities of two vertex sets respectively (\ie device-device and PHA-PHA similarities in this paper), and integrate them into a single objective function through shared embedding vectors. The rationale of using CSE as a baseline is to understand if high-order proximity matrix factorization itself can offer superior prediction results for the real world deployment.

\noindent \textbf{Excluded methods.} We also experimented with BiNE~\cite{gao2018bine}, an embedding model specialized for bipartite graphs and Jodie~\cite{kumar2019predicting}, a coupled recurrent neural network based embedding model. However, BiNE does not finish the computation for $DS_1$ after 72 hours while Jodie failed to train due to the out-of-memory issue in a TITAN V 12GB graphics card. Therefore, we exclude them from the comparison study for practicality reasons.

\noindent \textbf{Experimental Results}. 
Table~\ref{tab:exp_comparison} shows the prediction performance of \approach compared to the baseline systems. As it can be seen, \approach outperforms the baseline methods in all three datasets regarding all evaluation metrics. We can also note that the high-order proximity method (\ie CSE) performs better than first/second-order proximity methods, and first/second-order proximity methods perform better than preferential attachment. These results demonstrate that higher proximity between devices and PHAs is crucial to capture the implicit relations and improve prediction results. However, the high-order proximity baseline is less effective than \approach since it does not consider the diminishing coefficient with the increasing order (see Section~\ref{sec:embdding}). In general, \approach reports a TPR above 0.991 for a FPR of 0.0001 in all three datasets (\eg given $DS_1$ and 0.001 FPR, \approach is able to correctly predict 314,616 out 317,474 PHA installation events with 2,857 FNs), while the high-order proximity baseline only reaches a TPR of 0.893 in the best case (\ie given $DS_2$) for a FPR of 0.0001, representing a substantial 10\% TPR difference comparing to \approach (\ie high-order proximity leads to 283,504 TPs with 35,779 FNs which represent 10 times more FNs and 31,112 less TPs comparing to \approach). For different FPR values (0.0001 - 0.005), \approach consistently reports TPRs ranging from 0.991 to 0.998 for all three datasets. The high-order proximity baseline gets close to \approach's prediction performance given a FPR of 0.005. Yet, this represents a 50 times increase in FPR comparing to the optimal operational FPR of 0.0001. 
In addition, we can observe that the preferential attachment baseline performs the worst. This supports our hypothesis that popular PHAs are not the dominant factor when predicting future installations, but rather that rare PHAs also need to be factored in for effective predictions. 
Figure~\ref{fig:exp_comparison_auc} shows the ROC curve of \approach compared to the baseline systems. As it can be seen, \approach outperforms the baseline methods in all three datasets, which further confirms \approach's operational capability in the real world deployment (see Section~\ref{sec:exp_runtime} for \approach's runtime performance).

\subsection{Interpretation of \approach Prediction Performance (RQ2)}
\label{sec:exp_interpretation}

In this section, we aim at understanding why \approach performs better than the baseline methods (\eg ruling out the potential overfitting, etc). To this end, we leverage LargeVis~\cite{tang2016visualizing}, an efficient high-dimensional data visualization system, to project and visualize both positive (\ie observed) and negative (\ie non-existent) PHA installation events (\ie concatenated $2d$-dimensional vectors) into a 2-dimensional space. Such visualization can facilitate the straightforward description, exploration, and sense-making of the original data projected in the latent space by \approach (see Section~\ref{sec:methodology}). Our hypothesis is that, since we formulate the prediction task as a binary classification problem (\ie a PHA installation will be observed or not), \approach's low dimensional representations should be able to separate these two categories of events (\ie positive events should be apart from negative events) in the latent space. 

The LargeVis visualization results from all methods (see Section~\ref{sec:exp_comparison}) are shown in Figure~\ref{fig:exp_largevis}. For illustration purposes, we show the baseline results using $DS_1$ and \approach's results using all three datasets. Note that preferential attachment does not produce embeddings hence is excluded from the figure. As we can see, \approach can well separate positive (in yellow) from negative (in blue) PHA installation events (see Figure~\ref{fig:lagevis_approach}, Figure~\ref{fig:lagevis_approach_1w} and Figure~\ref{fig:lagevis_approach_1m}) in the latent space without relying upon the prediction models (see Section~\ref{sec:discussion}). At the same time, the high-order proximity method (Figure~\ref{fig:lagevis_cse}) can separate the two categories of events better than first/second-order proximity methods do (Figure~\ref{fig:largevis_lineo1} and~\ref{fig:largevis_lineo2}). This is consistent with the results obtained in the comparison study (see Section~\ref{sec:exp_comparison}). Our visualization results demonstrate that \approach can offer a better PHA installation event representation in the latent space, providing us with a good explanation of why \approach performs better than the baseline methods as shown in Section~\ref{sec:exp_comparison}. In Section~\ref{sec:discussion} we also show that the installation event representations obtained by \approach work well with different prediction models.

\subsection{Resilience to Data Latency (RQ3)}
\label{sec:exp_resilience}

\begin{table}[t]
\centering
\resizebox{\linewidth}{!} {
\begin{tabular}{|c|c|c|c|c|c|c|c|c|}
\hline
\textbf{Dataset} & \textbf{\begin{tabular}[c]{@{}c@{}}Training\\ ratio\end{tabular}} & \textbf{\begin{tabular}[c]{@{}c@{}}Data latency\\ ratio\end{tabular}} & \textbf{\begin{tabular}[c]{@{}c@{}}Test\\ ratio\end{tabular}} & \textbf{\begin{tabular}[c]{@{}c@{}}TPR\\ @\\ 0.0001\end{tabular}} & \textbf{\begin{tabular}[c]{@{}c@{}}TPR\\ @\\ 0.001\end{tabular}} & \textbf{\begin{tabular}[c]{@{}c@{}}TPR\\ @\\ 0.005\end{tabular}} & \textbf{\begin{tabular}[c]{@{}c@{}}ROC\\ AUC\end{tabular}} & \textbf{\begin{tabular}[c]{@{}c@{}}AP\end{tabular}} \\ \hline
\multirow{4}{*}{$DS_2$} 
 & 0.86 & 0.00  &  0.14 &  0.994 & 0.997  & 0.998 & 0.9994  & 0.9995  \\ \cline{2-9} 
 & 0.79 & 0.07  &  0.14 &  0.994 & 0.997  & 0.998 & 0.9994  & 0.9995  \\ \cline{2-9} 
 & 0.70  & 0.16  &  0.14  & 0.993 & 0.997 & 0.998 & 0.9994 & 0.9995  \\ \cline{2-9} 
 & 0.61 & 0.25  &  0.14 & 0.991 & 0.994  & 0.997 & 0.996 & 0.995 \\ \hline
\multirow{4}{*}{$DS_3$} 
 & 0.839 & 0.00  &  0.161 &  0.992 & 0.995  & 0.997 & 0.9994  & 0.9995  \\ \cline{2-9} 
& 0.769 & 0.07 & 0.161 & 0.992 & 0.995 & 0.997 & 0.9992 & 0.9994 \\ \cline{2-9} 
 & 0.679 & 0.16 & 0.161 & 0.991 & 0.994 & 0.995 & 0.9992 & 0.994 \\ \cline{2-9} 
 & 0.589 & 0.25 & 0.161 & 0.990 & 0.992 & 0.994 & 0.996 & 0.997 \\ \hline
\end{tabular}
}
\caption{\approach's resilience to data latency.}
\label{tab:exp_resilience}
\end{table}

The mobile security product from the major security company performs its detection periodically, and usually sends detection data to the backend when the devices are connected to a wireless network in order to minimize mobile data usage. From the data perspective, we observe the phenomenon that a device's detection data is less frequent and may be delayed for several hours or a couple of days. In light of this constraint, we expect \approach to cope with this limitation and remain operational. In this section, we look at how \approach performs when facing such data latency issue in a real world deployment.

\noindent \textbf{Experimental setup.} We use $DS_2$ and $DS_3$ (see Table~\ref{tab:exp_datasets}) to study \approach's prediction performance when dealing with the real world data latency issue. We randomly remove 7\%, 16\% and 25\% PHA installation events from the training data to simulate the data latency. Training, data latency and test ratio of these two datasets are summarized in Table~\ref{tab:exp_resilience}. We use the same parameters and the evaluation metrics laid out in Section~\ref{sec:exp_setup}. Note that we fix the test data ratio so that the prediction results with the simulated data latency are comparable to those without (see Table~\ref{tab:exp_comparison}).

\noindent \textbf{Experimental results.} Table~\ref{tab:exp_resilience} summarizes the results obtained using the aforementioned experimental setup. For clarification purposes, we also include the prediction results from Table~\ref{tab:exp_resilience} in Section~\ref{sec:exp_comparison}. Overall, \approach shows strong resilience to data latency. For example, \approach's prediction performance is not affected by any of the evaluation metrics when data latency ratio is 7\%. When data latency ratio increases to 16\%, \approach's prediction drops by 0.001 - 0.002 TPR given various FPR thresholds. Even when the data latency ratio reaches 25\%, \approach can still offer high accuracy with a small decrease of AUC and AP. For example, given 0.0001 FPR, \approach can reach 0.991 and 0.990 TPR in predicting PHA installations respectively for $DS_2$ and $DS_3$, dropping by 0.003 and 0.002 TPR respectively to those from full training data. These results confirm that the model can cope well with the data latency issue and predict PHA installations with high accuracy in a real world deployment. In Section~\ref{sec:discussion} we provide a detailed discussion on how \approach's resilience to data latency can deal with the potential adversarial attacks~\cite{bojchevski2019adversarial}.

\subsection{Reliability over Time (RQ4)}
\label{sec:exp_reliability}

\begin{figure}[t]
    \centering
    \includegraphics[width=0.7\linewidth]{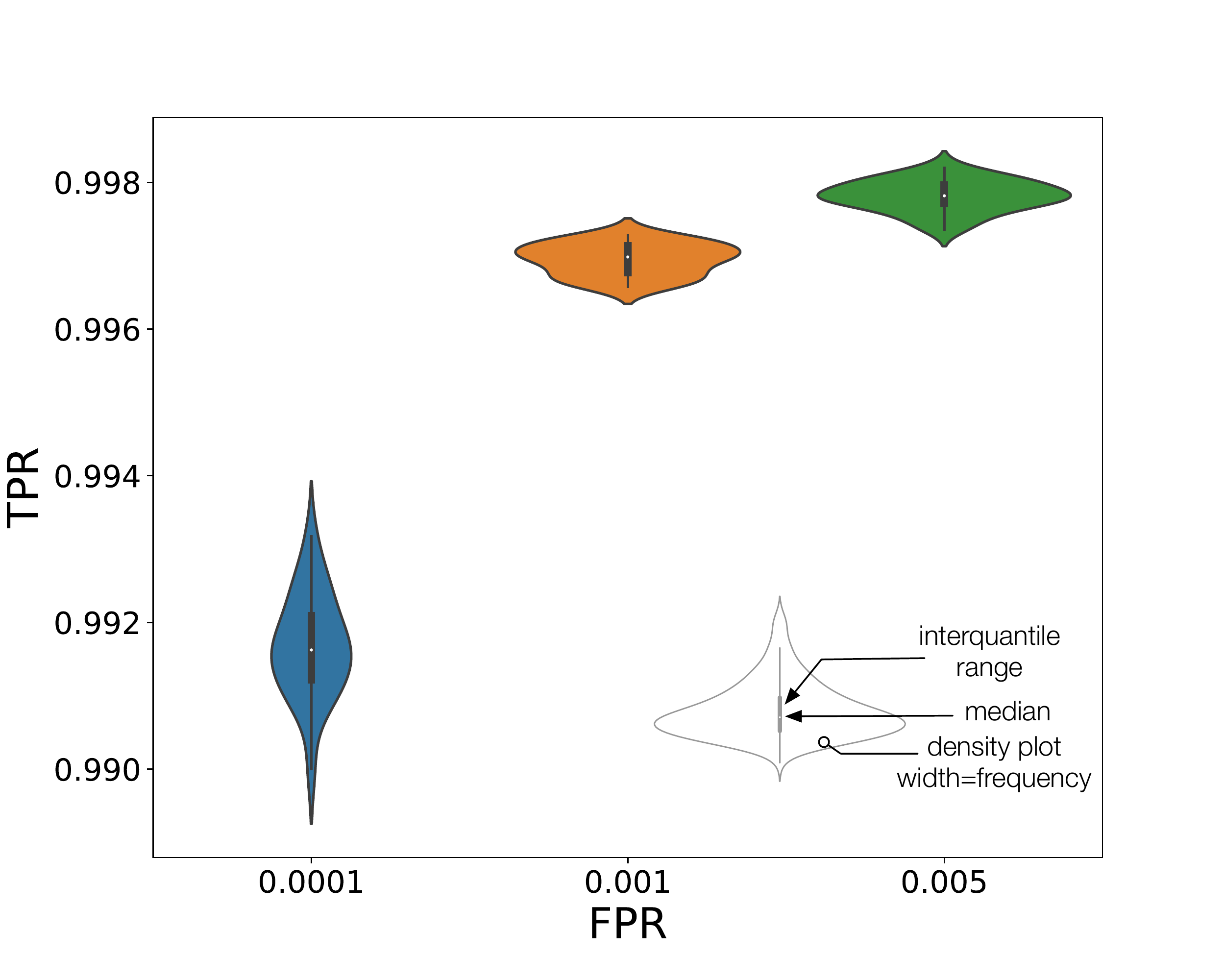}    
    \caption{Violin plot~\cite{hintze1998violin} for \approach's reliability study using a 7-day rolling window over $DS_3$. The white dot in the middle is the median value, the thick black bar in the centre represents the interquartile range and the contour represents distribution shape of the data. }
    \label{fig:reliability}
\end{figure}

In this section, we evaluate \approach's prediction accuracy when the system is deployed in the real world, where periodical retraining is required. Our goal is to evaluate the reliability of the model's prediction performance over time.

\noindent \textbf{Experimental setup.} We use $DS_3$ (see Table~\ref{tab:exp_datasets}) in this section to study prediction performance of \approach when periodical retraining is required. Training and test data are split in a rolling one-week window following the setup specified in Section~\ref{sec:exp_setup}. That is, we use the first 6 days in the window to form our training data and predict the PHA installations on the 7th day. This setup reflects the rapidly changing mobile threat landscape whereas new PHAs emerge daily. There are in total 24 evaluation points in this setup. We reuse the same parameters and the evaluation metrics as stated in Section~\ref{sec:exp_setup}.

\noindent \textbf{Experimental results.} We use violin plots to visualize the distribution of TPRs and its probability density per FPR threshold. Figure~\ref{fig:reliability} shows the results obtained using the aforementioned experimental setup. In this setup where retraining is required frequently, \approach's prediction performance remains steady. For example, given a 0.0001 FPR, \approach shows a 0.9916 TPR on average with a 0.0006 standard deviation. Given 0.001 and 0.005 FPRs, \approach shows a 0.997 TPR on average with a 0.0002 standard deviation and a 0.998 TPR with a 0.0002 standard deviation respectively. These results demonstrate that \approach can work well when frequent retraining is required, and underpins its real world practicability.

\subsection{Runtime Performance (RQ5)}
\label{sec:exp_runtime}

\begin{table}[t]
\centering
\resizebox{0.7\linewidth}{!} {
\begin{tabular}{|c|c|c|c|}
\hline
\textbf{Dataset} & \textbf{\begin{tabular}[c]{@{}c@{}}Graph \\ Construction\\ (seconds)\end{tabular}} & \textbf{\begin{tabular}[c]{@{}c@{}}\approach\\ runtime\\ (seconds)\end{tabular}} & \textbf{\begin{tabular}[c]{@{}c@{}} Pred. Eng.\\ training time\\ (seconds)\end{tabular}}  \\ \hline
$DS_1$ & 230.06 & 557.12 & 598.88 \\ \hline
$DS_2$ & 778.27 & 702.15  & 1,311.65 \\ \hline
$DS_3$ & 3,424.86 & 791.33 & 1,986.21 \\ \hline
\end{tabular}
}
\caption{\approach runtime performance}
\label{tab:exp_runtime}
\end{table}

Following our prediction performance analysis in the previous sections, we evaluate \approach's runtime performance to answer several key practical questions such as 1) \emph{how long does \approach take to construct a PHA installation graph}, 2) \emph{how long does \approach take to build low dimensional representations of PHA installations} and 3) \emph{how long does \approach take to train the prediction engine}. To answer these questions, we use all three datasets (see Table~\ref{tab:exp_datasets}). 
\approach's runtime performance is summarized in Table~\ref{tab:exp_runtime}. 

As we can see in Table~\ref{tab:exp_runtime}, all three components (see Figure~\ref{fig:architecture} in Section~\ref{sec:methodology}) of \approach scale linearly with regards to the input data size. In terms of graph reconstruction time, \approach takes 230.06 seconds to reconstruct the PHA installation graph given $DS_1$. Respectively, \approach takes 778.27 seconds given $DS_2$ (\ie one week data) and 3424.86 seconds given $DS_3$ (\ie one month data). This represents 3.5 and 14.9 times increase of time given roughly 6.7 and 29.2 times more raw input data. In terms of generating low dimensional edge representations, \approach takes 598.88 seconds to build the edge representations for $DS_1$. Respectively, \approach takes 702.15 and 791.33 seconds to build corresponding edge representations for $DS_2$ and $DS_3$. This represents 1.17 and 1.56 times increase of time given roughly 1.98 and 2.97 times increase of input edge sizes (see Table~\ref{tab:exp_datasets}). In terms of training the prediction engine, \approach also shows similar patterns. This implies that the end-to-end operation of \approach takes approximately 46.5 minutes for a typical usage scenario (\ie prediction for one week data using $DS_2$, see Section~\ref{sec:exp_reliability}). In summary, \approach's empirical linear scalability exemplifies its computational advantage and practical deployment in the wild.

\section{Case Studies}
\label{sec:case_study}

In this section, we provide two interesting case studies that we encountered while operating \approach. They demonstrate how \approach could be used as an early warning tool, and how it may explain the implicit connections between PHAs.

\subsection{Adware Installation Prediction}
\label{sec:case_study_adware}

\begin{figure}[t]
    \centering
    \includegraphics[width=0.9\linewidth]{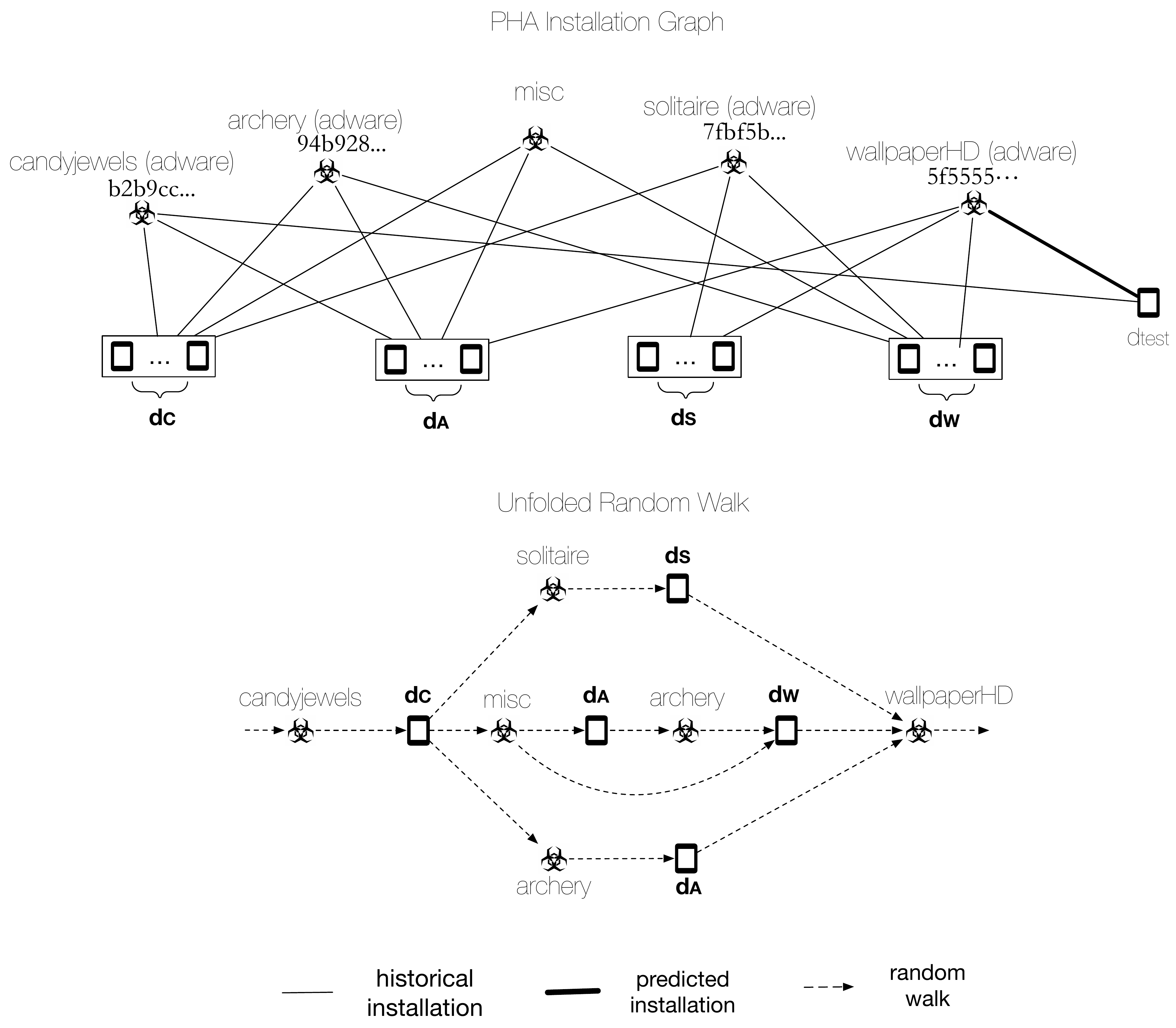}
    \caption{Case study: \approach predicts adware installation}
    \label{fig:case_study_adware}
\end{figure}

In this section, we carry out a detailed adware installation prediction case study. Figure~\ref{fig:case_study_adware} shows part of a PHA installation graph built by \approach on March 1, 2019. Vertices connected by thin solid lines are observed PHA installations and the thick solid line represents the prediction by \approach. There are four repackaged adware apps that aggressively use the MoPub library and consequently are not available from the Google Play store. Respectively they are \texttt{candyjewels} with SHA2 \texttt{B2B9CC...}, \texttt{archery} with SHA2 \texttt{94B928...}, \texttt{solitaire} with SHA2 \texttt{7FBF5B...} and \texttt{wallpaperHD} with SHA2 \texttt{5F5555...}. During the training, there are a number of random walks with various numbers of hits leading from \texttt{candyjewels} to \texttt{wallpaperHD}. 
The reason for this is likely that those apps were available on the same third party marketplace and suggested to populations of users with similar app installation history.
We only provide a snippet of the unfolded random walks (dash lines) in Figure~\ref{fig:case_study_adware} for clarification purposes. These random walks exhibit various  $l$-order proximities (see Section~\ref{sec:methodology}), and the accumulation of these walks enables \approach to capture the implicit relationship between \texttt{candyjewels} and \texttt{wallpaperHD}. At the test time, once a device $d_{test}$ installs \texttt{candyjewels}, \approach is able to predict that $d_{test}$ may install \texttt{wallpaperHD} with high confidence based upon the evidential information collected from these random walks. Following up with the prediction, \approach alerts the end user to be vigilant when installing \texttt{wallpaperHD} app by displaying a summarized description of the unfolded random walks. The end user can reach an informed decision to not install a PHA which might weaken Android's built-in security.

Additionally, we can observe that the unfolded random walks shown in Figure~\ref{fig:case_study_adware} reveal some interesting patterns that are not available in the detection data. For example, adware may explore a user's preference (\eg \texttt{candyjewels}, \texttt{archery}, and \texttt{solitaire} are all games), and lead the users to install additional adware (\eg \texttt{wallpaperHD}) which is favored by certain demographics group playing these games. \approach is able to capture these implicit patterns without collecting additional data and could warn targeted users ahead of time.

\subsection{Trojan Installation Prediction}
\label{sec:case_study_trojan}

\begin{figure}[t]
    \centering
    \includegraphics[width=\linewidth]{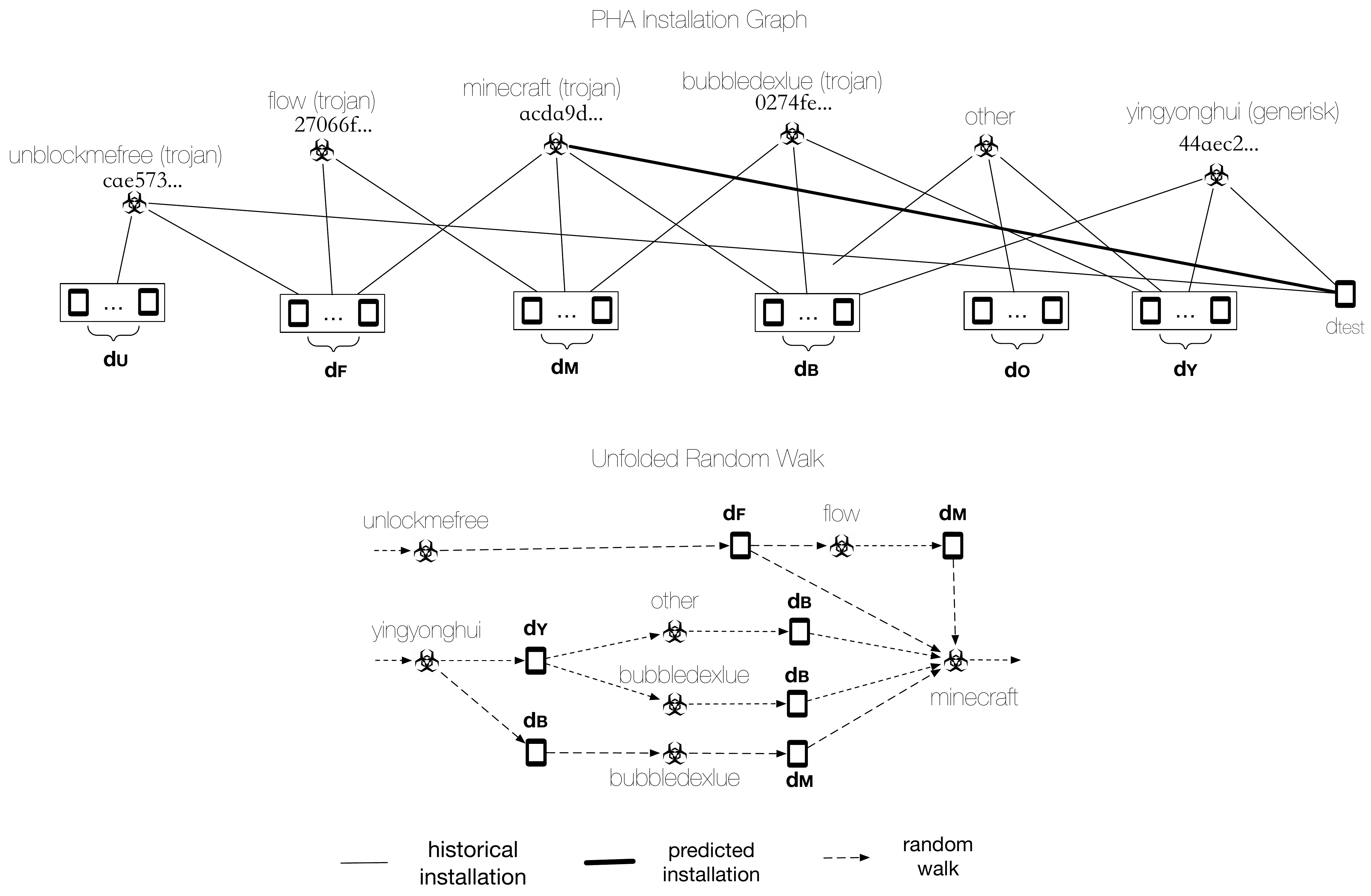}
    \caption{Case study: \approach predicts trojan installation}
    \label{fig:case_study_trojan}
\end{figure}

In this section, we carry out a detailed trojan installation prediction case study. Figure~\ref{fig:case_study_trojan} shows part of a PHA installation graph built by \approach on March 1, 2019. Thin solid lines represent observed PHA installations and the thick solid line represents the prediction by \approach. There are four trojans repackaged from popular games, respectively \texttt{unlockmefree} with SHA2 \texttt{CAE573...}, \texttt{flow} with SHA2 \texttt{27066F...}, \texttt{minecraft} with SHA2 \texttt{ACDA9D...} and \texttt{bubbledexlue} with SHA2 \texttt{0274FE...} from third party marketplaces.
In addition, we have a third party marketplace application \texttt{yingyonghui} with SHA2 \texttt{44AEC2...}. Note that \texttt{yingyonghui} functions as a gateway to applications, however it is itself associated with information leakage risk since it collects both network and SMS data without explicit notification to/consent from the users. The other applications with limited hits during the random surfing are classified as `\texttt{misc}'. We also provide a snippet of the unfolded random walks (dash lines) in Figure~\ref{fig:case_study_trojan} for illustration purpose as we did in Section~\ref{sec:case_study_adware}. As we can see in Figure~\ref{fig:case_study_trojan}, there are multiple sources from different application categories (\eg \texttt{unlockmefree} is a game and \texttt{yingyonghui} is a market application) that may lead to the installation of trojan application \texttt{minecraft}. For example, the trojan (\eg \texttt{minecraft}) makers can target third party marketplaces (\eg \texttt{yingyonghui}) which don't enforce code review, and leverage a user's inclinations to install this popular application for free. Additionally, the trojan makers can target specific application categories (\eg game), and tactically repackage popular titles (\eg \texttt{bubbledexlue},  \texttt{unlockmefree} etc.) and spread additional trojans. \approach is able to capture these different strategies by leveraging the collective PHA installation graph. At the test time, if a device $d_{test}$ installs \texttt{unlockmefree} and \texttt{yingyonghui}, \approach can predict the impending installation of the malicious version of \texttt{minecraft} with high confidence.

\section{Discussion}
\label{sec:discussion}

\begin{table*}[t]
\centering
\resizebox{0.6\linewidth}{!} {
\begin{tabular}{|c|c|c|c|c|c|c|c|c|c|c|}
\hline

\multirow{2}{*}{Method} & \multicolumn{5}{c|}{Logistic Regression} & \multicolumn{5}{c|}{XGBOOST} \\ \cline{2-11}
 & \begin{tabular}[c]{@{}c@{}}TPR\\ @\\ 0.0001\end{tabular} & \begin{tabular}[c]{@{}c@{}}TPR\\ @\\ 0.001\end{tabular} & \begin{tabular}[c]{@{}c@{}}TPR\\ @\\ 0.005\end{tabular} & \begin{tabular}[c]{@{}c@{}}ROC \\ AUC\end{tabular} & AP & \begin{tabular}[c]{@{}c@{}}TPR\\ @\\ 0.0001\end{tabular} & \begin{tabular}[c]{@{}c@{}}TPR\\ @\\ 0.001\end{tabular} & \begin{tabular}[c]{@{}c@{}}TPR\\ @\\ 0.005\end{tabular} & \begin{tabular}[c]{@{}c@{}}ROC\\ AUC\end{tabular} & AP \\ \hline
1st-order prox. & 0.151 & 0.172 & 0.188 & 0.446 & 0.551 & 0.500 & 0.552 & 0.603 & 0.894 & 0.914 \\ \hline
2nd-order prox. & 0.722 & 0.821 & 0.869 & 0.923 & 0.959 & 0.633 & 0.768 & 0.852 & 0.981 & 0.982 \\ \hline
high-order prox. & 0.782 & 0.832 & 0.875 & 0.96 & 0.97 & 0.684 & 0.850 & 0.901 & 0.991 & 0.993 \\ \hline
\approach & \textbf{0.900} & \textbf{0.903} & \textbf{0.98} & \textbf{0.97} & \textbf{0.959} & \textbf{0.917} & \textbf{0.947} & \textbf{0.982} & \textbf{0.999} & \textbf{0.999} \\ \hline
\end{tabular}
}
\caption{\approach's prediction performance (using $DS_1$) with alternative prediction models, respectively logistic regression and gradient boosting. }
\label{tab:other_classifiers}
\end{table*}

\begin{table*}[t]
\centering
\resizebox{0.5\linewidth}{!} {
\begin{tabular}{|c|c|c|c|c|c|c|}
\hline
\textbf{method} & \textbf{operation} & \textbf{\begin{tabular}[c]{@{}c@{}}TPR\\ @\\ 0.0001\end{tabular}} & \textbf{\begin{tabular}[c]{@{}c@{}}TPR\\ @\\ 0.001\end{tabular}} & \textbf{\begin{tabular}[c]{@{}c@{}}TPR\\ @\\ 0.005\end{tabular}} & \textbf{AUC} & \textbf{AP} \\ \hline
Average & $(a+b)/2$ & 0.045 & 0.191 & 0.402  &  0.926 & 0.930 \\ \hline
Hadamard & $(a_1 * b_1,..., a_i*b_i)$ & 0.056  & 0.159 & 0.347 & 0.921 & 0.922 \\ \hline
weighted L1 & $(|a_1-b_1|, ..., |a_i-b_i|)$ & 0.053  & 0.187 & 0.347 & 0.896 & 0.903 \\ \hline
weighted L2 & $((a_1-b_1)^2, ..., (a_i-b_i)^2)$ & 0.052 & 0.184  & 0.347 & 0.895 & 0.903 \\ \hline
concat & $[a_1, ..., a_i, b_1, ..., b_i]$ &\textbf{0.991} & \textbf{0.996} & \textbf{0.998} & \textbf{0.999} & \textbf{0.999}   \\ \hline
\end{tabular}
}
\caption{\approach's prediction performance (using $DS_1$) with alternative edge representations.}
\label{tab:other_edge_represenations}
\end{table*}

\noindent \textbf{Influence of different prediction models.} In Section~\ref{sec:exp_interpretation} we demonstrate that \approach can well separate positive (\ie observed) from negative (\ie non-existent) PHA installation events in the latent space without relying on the prediction model (see Section~\ref{sec:exp_setup}). For discussion purposes, we also illustrate how different prediction models may impact evaluation results. To this end, we select logistic regression~\cite{hosmer2013applied} and gradient boosting~\cite{natekin2013gradient}, respectively a linear and an ensemble model, to demonstrate the impact of alternative prediction models. As we can see in Table~\ref{tab:other_classifiers}, \approach is less affected by different models and attains reasonable prediction capability in both cases. For example, \approach is able to offer useful prediction performance (\eg 0.900 TPR and 0.917 TPR given 0.0001 FPR respectively using logistic regression and gradient boosting machine). In contrast, all the baseline methods have shown substantial drops in terms of prediction accuracy. This gives us additional confidence that \approach can offer a better PHA installation event representation and consequent prediction in the latent space.\\

\noindent \textbf{Influence of different edge representations.} The core of \approach is a graph representation learning algorithm that learns the low dimensional vertex representations in the latent space. Other than the concatenation used in this paper, there are alternative methods to combine two vertex representations to form a PHA installation event representation. To this end, we used four other methods, respectively \emph{average}, \emph{Hadamard}, \emph{weighted L1} and \emph{weighted L2}, to combine two vertex representations. We train \approach using these edge-wise representations obtained with the methods shown in Table~\ref{tab:other_edge_represenations}.  As we can see in Table~\ref{tab:other_edge_represenations}, all four alternative PHA installation event representation methods do not offer comparable prediction performance. \approach with the concatenation of vectors is consistently the best edge representation.  \\

\noindent \textbf{Evasion.} 
\approach may be subject to evasion by attackers.
At its core, \approach relies on graph representation learning to embed both device and PHA vertices into a latent space. 
Therefore, it may be subject to adversarial attacks~\cite{bojchevski2019adversarial} from an adversary who might influence the representation learning by adding/removing edges from the PHA installation graph. 
In terms of adding edges to influence prediction results, one technique that could be used by the adversaries is leveraging the strategy used by mimicry attacks, \ie evading prediction by injecting many irrelevant applications into alternative markets and ensure the users to install them to cover the prediction of a real trojan application.  
We argue that monetization is a key factor for mobile PHA makers. 
If they could install irrelevant applications in mobile devices, they would easily monetize from these applications. 
In other words, it would not make sense for the adversaries to distribute irrelevant applications since it would make PHA delivery much more expensive to prevent \approach from correctly predicting future installations. 
In terms of removing edges to influence prediction results, it is infeasible from a practicality perspective. 
Once a PHA installation is observed by the security product from this security company, the adversaries cannot remove the edges (\ie install events) from the PHA installation graph. One possible strategy can be leveraged by the adversaries is to prevent the security product from sending back the telemetry data.
We already show in Section~\ref{sec:exp_resilience} that \approach can well cope with data latency and predict PHA installations with high accuracy in the real world deployment.  
Additionally, an adversary may inject multiple similar PHAs slightly differing from each other. 
The PHAs would target a similar audience and their installation would be predicted with similar probabilities. 
From a performance perspective, such strategy would reduce \approach's TPR since the users will not install all the very similar PHAs. 
From a practical perspective, the users will be wrongly notified about a PHA not fitting their intention. 
This could consequently lead to an increased ``customer churn rate.''
One plausible approach to mitigate this evasion attack is using in-app messages to ask the end user to be vigilant of the markets and the developers where they may install apps from (\eg using official markets).
The rationale of using in-app messages is to explain to the end user the potential risk associated with the known on-device PHAs.
More importantly, the end users would have contextual information (\eg which PHAs in what category) that may be more effective to raise their awareness as observed by~\cite{micallef2017stop}.
This way, the end user would benefit from an informative explanation rather than a short notification via the status bar.
Consequently, the device’s owner can still be warned about the threat and reach an informed decision to not install a PHA from that category which might weaken their device security. \\

\noindent \textbf{Bias.}
Our dataset is biased towards the end users of a single mobile security product, and therefore still presents some biases.
However, the distribution of devices used in this study is not heavily skewed towards any specific region.
It is also challenging to ascertain the coverage of PHAs covered by our study since it is infeasible to determine the total number of all PHAs.
\approach, however, is retrained frequently to accommodate newly discovered PHAs.
That is, \approach accumulates its knowledge of known PHAs when it is retrained. 
For instance, if a PHA was not initially included to train \approach by the security company, this PHA will later be used to retrain \approach once its signature and associated detection events are made available to the security company. 
As such, we believe the bias initially introduced by \approach will diminish over time.

\noindent \textbf{Limitations.} As any prediction system, \approach has some limitations. The main limitation of \approach
is that it can only predict PHAs that are already known to be malicious. 
This makes sense in our application case, since our goal is to solve the current limitations of the Android security model, which allow PHAs to act undetected on their victim devices until the user decides to uninstall them.
One way was to reduce this limitation is to retrain \approach frequently, as new PHAs become known. 
As we showed in Section~\ref{sec:exp_reliability}, \approach can work well on different observation window under frequent retraining, underpinning its real world practicability. 
We argue that the advantage of proactive moderation compared to existing Android anti-malware solutions significantly overcomes these limitations.

\section{Related Work}
\label{sec:related_work}

\noindent \textbf{Predictive security.} 
Prior work studied the feasibility of forecasting future security incidents using observed historical information~\cite{sun2018data}.
The core idea of these research work is to extract a set of pre-defined features from historical data, and train a machine learning model to predict, in a binary format, if a vulnerability is likely to be exploited~\cite{bozorgi2010beyond}, if a PoC of a vulnerability would be devised in the real world~\cite{sabottke2015vulnerability}, if an enterprise would be breached using publicly available security incident datasets~\cite{liu2015cloudy}, the volume of actual malware infections in a country~\cite{kang2016ensemble}, the likelihood of endpoints at risk of infection in the future~\cite{bilge2017riskteller}, etc. 
In recent years, researchers also leveraged deep learning techniques such as RNN~\cite{shen2018tiresias} and CNN~\cite{sharif2018predicting} to predict the \emph{exact} upcoming security events. 
Our approach focuses on learning low dimensional representations of global malware installation graph and predict emerging malware installations. \\

\noindent \textbf{Applications of representation learning in security research.} 
The core idea of these research work is to captures the posterior distribution of the underlying explanatory factors for the observed input in latent space. 
Lin~\etal~\cite{lin2017poster,lin2018cross} propose a function representation learning method to obtain the high-level and generalizable function representations from the abstract syntax tree (AST). 
Rhodeb~\etal~\cite{rhode2018early} use RNNs to learn the representations of their behavior profiles. 
Ding~\etal~\cite{ding2019asm2vec} represents an assembly function as a control flow graph (CFG) and leveraging a customized PV-DM model~\cite{le2014distributed} to learn the latent representations of assembly functions.  
Shen~\etal~\cite{shen2019attack2vec} propose ATTACK2VEC model to understand the evolution of cyberattacks in latent space. \\

\noindent \textbf{Graph-based malware detection.} 
The core idea of graph-based malware detection is modeling the interactions among malware, endpoints and network servers as graphs, and leverage various machine learning models to understand the patterns and detect previous unknown malicious files or activities.  
For example, CAMP~\cite{rajab2013camp}, Mastino~\cite{rahbarinia2016real}, and Polonium~\cite{nachenberg2011polonium} built graphs from binary activity data and detect malware. 
Similarly, Marmite~\cite{stringhini2017marmite}, NAZCA~\cite{invernizzi2014Nazca}, AESOP~\cite{tamersoy2014guilt} and Kwon~\etal~\cite{kwon2015dropper} built graphs from binary download/distribution data and detect previous unknown malware.  
Different from these approaches, \approach predicts PHA installations in the future instead of detecting them. \\

\noindent \textbf{Closest work.} 
The closest work to this paper is highly predictive blacklisting (HPB)~\cite{zhang2008highly}. 
Essentially, HPB builds a victim correlation graph using Jaccard similarity over occurrences of overlapping attackers among victims. 
However, HPB cannot scale to large dataset due to its quadratic time complexity incurred by Jaccard similarity. 
From a theoretical perspective, HPB is equivalent to weighted one mode projection of a bipartite graph~\cite{zhou2007bipartite} which only considers explicit attack-victim relationship. 
In contrast to HPB, we demonstrate that \approach has an empirical linear scalability coping with millions of vertices, and considers both explicit and implicit relationships offering accurate PHA installation predictions. 

\section{Conclusion}
\label{sec:conclusion}

We presented \approach, a system that learns latent relationships between PHAs and mobile devices and leverages them for prediction. 
We showed that it is possible to predict PHA installation on mobile devices with very accurate results (0.994 TPR with 0.0001 FPR) up to one week ahead of the real installations, and our system is performant and can account for delays in receiving data, which are typical in real world deployments.
This prediction approach has the potential of successfully complementing current on-device anti-malware systems, which are inherently reactive. 
We plan to study how to devise effective warning that could pro-actively sway users into not installing the PHAs that they will encounter, for example while browsing third party marketplaces.
To this end, we plan to work in the spirit of past research in usable security and warning design~\cite{egelman2008you,thompson2019web}.

\section*{Acknowledgments}

We wish to thank the anonymous shepherd and reviewers for their feedback in improving this paper.

\bibliographystyle{plain}
\bibliography{mobile}

\end{document}